\documentstyle[12pt]{article}
\catcode`\@=11
\@addtoreset{equation}{section}

\def\@normalsize{\@setsize\normalsize{15pt}\xiipt\@xiipt
\abovedisplayskip 14pt plus3pt minus3pt%
\belowdisplayskip \abovedisplayskip
\abovedisplayshortskip  \z@ plus3pt%
\belowdisplayshortskip  7pt plus3.5pt minus0pt}
\def\small{\@setsize\small{13.6pt}\xipt\@xipt
\abovedisplayskip 16pt plus3pt minus3pt%
\belowdisplayskip \abovedisplayskip
\abovedisplayshortskip  \z@ plus3pt%
\belowdisplayshortskip  7pt plus3.5pt minus0pt
\def\@listi{\parsep 4.5pt plus 2pt minus 1pt
            \itemsep \parsep
            \topsep 9pt plus 3pt minus 3pt}}
\def\underline#1{\relax\ifmmode\@@underline#1\else
	$\@@underline{\hbox{#1}}$\relax\fi}
\@twosidetrue
\catcode`@=12
\catcode`\@=11

\def\FERMIPUB{}
\def\FERMILABPub#1{\def\FERMIPUB{#1}}
\def\ps@headings{\def\@oddfoot{}\def\@evenfoot{}
\def\@oddhead{\hbox{}\hfill
	\makebox[.5\textwidth]{\raggedright\ignorespaces --\thepage{}--
	\hfill {\rm FERMILAB--Pub--\FERMIPUB}}}
\def\@evenhead{\@oddhead}
\def\subsectionmark##1{\markboth{##1}{}}
}
\ps@headings
\catcode`\@=12
\relax

\newskip\humongous \humongous=0pt plus 1000pt minus 1000pt

\newif\ifdtup

\def\oldreffmt#1{\rlap{[#1]} \hbox to 2\parindent{}}

\def\figfmt#1{\rlap{Figure {#1}} \hbox to 1in{}}

\def\ie{\hbox{\it i.e.}{}}	
\def\eg{\hbox{\it e.g.}{}}	
\def\etal{\hbox{\it et al.}}

\def\tr{\mathop{\rm tr}}
\def\Tr{\mathop{\rm Tr}}
\def\Im{\mathop{\rm Im}}

\def\bra#1{\left\langle #1\right|}
\def\ket#1{\left| #1\right\rangle}
\def\VEV#1{\left\langle #1\right\rangle}

\def\slash#1{#1\!\!\!/\!\,\,}
\def\beq{\begin{equation}}
\def\eeq{\end{equation}}
\def\bea{\begin{eqnarray}}
\def\eea{\end{eqnarray}}
\def\half{\frac{1}{2}}

\def\bq{\begin{quote}}
\def\eq{\end{quote}}

\def\half{\frac{1}{2}}
\def \etal {{\it et al.}\ }
\relax
\begin{document}
\par \vskip .05in
\FERMILABPub{93/059--T}
\begin{titlepage}
\begin{flushright}
SSCL--PP--243\\
FERMI--PUB--93/059--T\\
April 15, 1993
\end{flushright}
\vfill
\begin{center}
{\large \bf Chiral Dynamics and  Heavy Quark Symmetry\\
    in a Solvable Toy Field Theoretic Model }
 \end{center}
  \par \vskip .1in \noindent
\begin{center}
{\bf William A. Bardeen}
  \par \vskip .02in \noindent
{Theoretical Physics, SSC Laboratory \\
2550  Beckleymeade Ave., Dallas, Texas 75237--3946;\\
(Bardeen@SSCVX1)}
  \par \vskip .02in \noindent
{ and\\}
  \par \vskip .02in \noindent
 {\bf Christopher T. Hill }
  \par \vskip .02in \noindent
{Fermi National Accelerator Laboratory\\
P.O. Box 500, Batavia, Illinois, 60510\\
(Hill@FNAL)}
  \par \vskip .02in \noindent
\end{center}
\begin{center}{\large Abstract}\end{center}
\par \vskip .01in
\begin{quote}
We study a solvable
QCD--like toy theory, a generalization
of the Nambu--Jona-Lasinio model, which implements
chiral symmetries of light quarks and heavy
quark symmetry. The chiral symmetric and chiral broken phases can be
dynamically tuned.  This implies a parity doubled heavy--light meson system,
corresponding to a
$(0^-,1^-)$ multiplet and a $(0^+,1^+)$ heavy spin
multiplet.  Consequently
the mass difference of the two multiplets  is given by a Goldberger--Treiman
relation and $g_A$ is found to be small.   The  Isgur--Wise function,
$\xi(w)$, the decay constant,  $f_B$, and other observables are studied.
\end{quote}
\vfill
\end{titlepage}

\noindent
{\bf I. Introduction}
\vskip .1in
\noindent
In recent years, QCD applied to systems containing a single very massive
quark, where one can imagine the limit $M\rightarrow \infty$ to be a reasonable
physical approximation,
has been the subject of considerable attention [1-5].
The pseudoscalar and vector mesons
containing one very massive and one light quark become degenerate
in the $M\rightarrow \infty$ limit, due to a heavy quark spin
symmetry again valid to $1/M$.  Moreover, Isgur and Wise
[1] pointed out that transition amplitudes, such as weak decays, involving
heavy quarks
are described by a flavor independent function of the invariant difference
in  4--velocities, $\xi(v'\cdot v)$, and therefore a
heavy quark spin--flavor symmetry, $SU(2N_f)$ exists, valid to order $1/M$.
Georgi
has given a useful field theoretic construction of this limit [5],
and has studied the consequences and phenomenological applications
of the theory, such as the computation of the QCD anomalous dimension
which controls the perturbative evolution of $\xi(v'\cdot v)$ with scale
for $v'\cdot v < 1$.

For many purposes  one must also implement the
chiral symmetries of the light quarks,
in addition to the heavy quark symmetry.  The heavy quark (HQ)
and chiral light quark (LQ) symmetries together control the interactions
of heavy--light (HL) mesons with pions and $K$--mesons, etc.
Several authors have written down model independent
chiral Lagrangians which
involve these symmetries at the meson level  [6-12]. A number of
studies of the phenomenological applications of these chiral
Lagrangians have been undertaken, such as  the computation
of the chiral log radiative corrections to $\xi(v'\cdot v)$, [7]
associated with $SU(3)\times SU(3)$ breaking terms, the study of radiative
and meson decays of heavy mesons [11], and chiral dynamics
including the effects of excited heavy mesons [12].

The chiral Lagrangian introduced by Wise [6] represents a straightforward
implementation of the heavy quark and light flavor symmetries
in the nonlinear {\em current form}. In this form one need only
identify the linear flavor symmetries, like isospin or
$SU(3)$, and the chiral effective Lagrangian, to leading
order in the momentum expansion, is automatically
determined, up to an unspecified axial
vector coupling constant $g_A$. This effective
Lagrangian is then manifestly invariant under the
usual global flavor symmetries, and the full set of chiral
transformations are local gauge--like transformations
which are functionals of the pions.  The underlying chiral representations
of the heavy mesons need never be specified.
Model independent approaches are clearly the most
reliable way in which to minimally implement the physical symmetries.

We may wish, however, to go closer to the underlying chiral dynamics
than the model independent approaches allow.
We may pose additional questions within dynamical models which can
reveal additional physical consequences to the real world.
For example, is there a
more primitive {\em chiral form} of the Lagrangian in which the explict
chiral representations of the heavy mesons are identified? A related
question in the broken phase is: what
is the analogue of the Goldberger--Treiman relation in the heavy meson system,
i.e.,  what receives mass from the chiral condensate's mass gap?
In the case of the nucleon--meson system we can similarly write
the chiral Lagrangian in the nonlinear current form, never
having to specify the precise
chiral representations of nucleons.  However, if we ask
for the {\em linear} chiral form we also know
the answer:  the left--handed (right--handed) nucleon is
assigned to a $(0,\half)$, ($(\half, 0)$) representation under
$SU(2)_R\times SU(2)_L$. Most of the nucleon mass arises from
the chiral condensate, or the VEV of $\Sigma$ which is
$(\half,\half)$.  We know this because the Goldberger--Treiman relation yields
the pion--nucleon coupling constant in terms of the nucleon mass
$g_{NN\pi}\approx m_N/f_\pi$, and $G_A\approx 1$.

In the case of heavy mesons, however, it is clear that the meson mass arises
primarily from the mass of the heavy constituent quark, such as the
$b$--quark,
and  the chiral mass gap is  a perturbation. Our question then is
related to the outcome of a gedanken experiment: what happens
to the heavy--light meson spectrum if we could somehow restore
the chiral symmetry, maintaining the other features of confining
QCD? While
the nucleon mass goes to zero in this gedanken limit, leaving degenerate
(approximately) massless left-- and right--handed states, the heavy
meson masses must remain (approximately) unaffected. Yet, the explicit
linear chiral symmetry $SU(2)_L\times SU(2)_R$ must somehow be realized in
the heavy meson mass spectrum in this limit.
This leads to the conclusion that the
ground--state must become doubly degenerate with even and odd parity
mesons $B_1$ and $B_2$ respectively, and these must form
the $SU(2)_L\times SU(2)_R$
representations in the linear  combinations $(B_1\pm B_2)/\sqrt{2}$.
Therefore, the breaking of the chiral symmetry  leads to a mass gap
between these parity partners and associated pionic transitions
between parity partners will occur (at the end of Appendix B we give a
schematic discussion of parity doubling).

It is difficult to imagine that a simple potential model can
capture this phenomenon.  The chiral symmetry limit is
relativistic, and the chiral symmetry breaking is a dynamical
rearrangement of the vacuum.  Thus, the naive picture of a heavy meson as
a boundstate of a heavy quark and a constituent quark will miss
those aspects of the physics which involve the necessary mixing
of the parity doubled states. This will show up in the present
analysis in the meaning and quantitative estimate
of $g_A$, and the analogue to the Goldberger--Treiman
relation.

  Thus, to better understand these issues it is interesting,
if not essential,
to study simple, solvable, strongly coupled toy
field--theoretic models in which both
heavy quark and chiral symmetries
are present at the fundamental quark level, and the dynamics of chiral
symmetry breaking is made explicit. We consider presently
the simplest such scheme.  We emphasize at the outset
that this toy model is unrealistic  and is
intended only to convey the schematics of
QCD chiral dynamics in heavy--light mesons
(although we will brazenly attempt a fit to data).
The simple model we consider is based upon
a local gluonic current--current interaction Lagrangian:
\beq
-\frac{g^2}{\Lambda^2}\sum_i\overline{\psi}_i
\gamma_\mu\frac{\lambda^A}{2}\psi_i
\sum_j\overline{\psi}_j
\gamma^\mu\frac{\lambda^A}{2}\psi_j
\eeq
where the $(i,j)$ sums extend over all of the fundamental fermion flavors, both
heavy and light, and we sum over the octet color index $A$.
We view eq.(1) as
essentially a QCD--inspired generalization of the Nambu--Jona-Lasinio model.
For small $g$ eq.(1) corresponds to  the  low--energy
perturbative interaction generated by
the exchange of a  ``massive gluon'' of mass $\Lambda/\sqrt{2}$. We propose
to study this model  using the technique
of the large--$N$ expansion, or equivalently, the fermion bubble approximation,
with a cut--off at $\Lambda$.  The model is exactly solvable in leading order.

Solving the theory
in the leading large--$N$ approximation  is equivalent to factorizing
eq.(1) into auxiliary fields describing the   composite pions and
heavy--light mesons, at the scale $\Lambda$ and integrating out the heavy and
light quarks to
generate the effective Lagrangian at a scale $\mu <\Lambda$.
The hadrons in our model, both
light and heavy, appear as dynamically generated boundstates.
In the light quark, meson sector we recover the chiral quark model
of Manohar and Georgi [13] (with $g_A^q=1$).  In the heavy meson sector we
produce
various boundstates of the heavy quark and the light  quarks,
and the full effective Lagrangian of these heavy meson boundstates coupled to
light mesons is determined.  The effective Lagrangian is manifestly heavy
quark--spin and chirally symmetric.

We will make certain further simplifying assumptions, keeping only
terms in the renormalized effective
Lagrangian that are $\sim O(1)$ or $\sim O(\mu \ln(\Lambda/\mu)/\Lambda)$,
while dropping subleading terms $\sim O(\mu/\Lambda)$.
We emphasize that we have in mind, presently,
a hierarchy of scales, $\mu < \Lambda <<M$,
where $M$ is the heavy quark mass scale.  The
momentum--space loop integrals will extend from $\mu$ to
the cut--off $\Lambda$.  We view in the context of the model
$\Lambda $ to be a physical scale below which the
theory is nonperturbative in $g$, but above
which an effective softening of the point--like approximation due to the
perturbative $1/q^2$ gluon propagator
takes place.  While it is tempting to identify $\Lambda$
with $\sim \Lambda_{QCD}$, we
would hope that $\Lambda \sim 1 $ GeV emerges from a fit
to the physical quantities derived in the model.  In fact, the
simplest attempt at a fit
to $f_B$ and $f_\pi$ yields  $\Lambda
\sim 1.35$ GeV, and most of the light sector observables
are obtained within a factor of two.  $\mu$ is an infrared cut--off
which we would like to identify with the scale of light
constituent quark masses.

In the unbroken chiral symmetry phase the model
produces the necessary degenerate parity doubling of
the threshold spectrum of heavy mesons.
In addition to the usual pseudo--scalar
and vector HL mesons (the $B$ and $B^*$ mesons
which form a $(0^-,1^-)$ heavy quark symmetry multiplet),
there is  necessarily a scalar and pseudo--vector
HL meson boundstate generated, which is a consequence of the
chiral symmetry. This is identified with the $s_l^{\pi_l}= \half^-$
p--wave radially excited mesons (in the D--system
this is distinct from the
observed $(D_2)$ states which are $s_l^{\pi_l}= \frac{3}{2}^-$,
(see Ming--Lu et.al. [14]).
Unfortunately, these states have not been observed and will be
fairly broad resonances, but they may ultimately be detectable.
Technically, in the symmetric phase we must hold $\mu$ fixed
at a nonzero value to protect from infrared singularities.

While the HQ symmetry maintains the degeneracy within the $(0^-, 1^-)$ and
$(0^+, 1^+)$ multiplets,
unbroken chiral symmetry implies the degeneracy of the two multiplets
themselves.
As we vary the model's
coupling parameter to dynamically
induce the chiral symmetry breaking,
the theory develops a mass gap.  This leads to a calculable
mass splitting, elevating the $(0^+, 1^+)$ HQ multiplet and depressing the
$(0^-, 1^-)$.
 The mass gap between the groundstate
mesons and the resonances is constant in the $M\rightarrow \infty$
limit and is given essentially by $\sim {g}f_\pi$.
{\em This is the analogue of the  Goldberger--Treiman relation of the theory,}
and
is probably more general than our specific toy model result.
Moreover, as a general result of the parity doubling, the axial vector
coupling constant $g_A$ is not necessarily expected to be close to unity.
In fact, $g_A$
tends to be small based upon our fit, $\approx 0.32$
(see Appendix B(ii);
it occurs here as a term of order $\ln(\Lambda/\mu)/\Lambda$,
which is subleading to $1$).
This is a prediction which is thus
far consistent with the upper limit
in processes like $D^*\rightarrow D +\pi$,
though a measurement of the full $D^*$ width
is still lacking to date.
In the limit of
very low $q^2$ pion emission we can decouple the heavier parity
doubling states to return to the effective chiral Lagrangian
for the $(0^-, 1^-)$ groundstate mesons of ref.[6]. There
remain in the low energy effective
Lagrangian potentially important effects of the heavy resonances
in chiral perturbation theory [12].

Thus, a key
result we find is that the chiral mass gap, and hence
an analogue  Goldberger--Treiman relation, refers to the
{\em splitting between parity conjugate heavy meson multiplets} in a
heavy--light
meson theory, i.e., {\em heavy
meson chiral theory is a parity doubled implementation of
chiral symmetry}. There are many other issues of the
applicability of the chiral theory and its consequences which
the present analysis will attempt  to address.
Though  not entirely realistic,
the model is completely solvable for various observables.
$f_B$ is determined in terms of the short distance
cut--off on the theory
and  the Isgur--Wise function is
computed.  The Isgur--Wise function result in the
present model involves
issues of going beyond the chiral logs, which arise
also in matching composite mesons onto QCD.
 We will discuss this issue which is related
to consideration of reparameterization invariance [14-16].

\vskip .1in
\noindent
{\bf II. Toy Model Field Theory with Chiral and Heavy Quark Symmetry}
\vskip .05in
\noindent
{\em (i) The Light Quark Chiral Dynamics }
\vskip .1in
Let us write the effective Lagrangian in the light quark sector,
including the current--current form of the
light fermion interaction Lagrangian of eq.(1):
\beq
{\cal{L}} = \overline{\psi}(i\slash{\partial} - m_q )\psi
-\frac{g^2}{\Lambda^2}\overline{\psi}
\gamma_\mu\frac{\lambda^A}{2}\psi\overline{\psi}
\gamma^\mu\frac{\lambda^A}{2}\psi
\eeq
For concreteness we will take $\psi = (u, d)$,
$\lambda^A$ are color matrices, and in
the limit that the quark mass matrix $m_q\rightarrow 0$, we
have an exact chiral $SU(2)_L\times SU(2)_R$ invariant
Lagrangian. This is a single  gluon exchange potential,
generated by a fake, massive gluon of mass $\Lambda/\sqrt{2}$.  We treat the
physics on scales $q^2<\Lambda^2$ using eq.(2), in a
fermion bubble approximation, imposing a UV loop
momentum cut--off of  $\Lambda$.    Well above the
scale $\Lambda$ we would imagine the potential to soften into a $1/q^2$
perturbative gluon exchange, hence $\Lambda $ plays the role
of a matching scale between strong infra--red physics and
weak ultraviolet QCD.
Finally, the ``theory'' consists of integrating out
the fermions  down to an infrared scale $\mu$, keeping induced
terms of order $\Lambda^2$, $\Lambda$ and $\ln(\Lambda/\mu)$
(we will discard terms that are finite, thus subleading, in the infinite
$\Lambda$ limit as a simplifying approximation).
This generates an effective Lagrangian of composite particles.
This is our essential approximation
to the infra--red strong coupling behavior of QCD, or
the ``brown muck'' of heavy--light physics.
Overall, this is certainly a drastic approximation.  Truncating on $dim=6$
operators is, in a sense, a pure $s$--wave approximation
to QCD, and cannot dynamically confine the quarks
and discarding
the subleading terms will limit the quantitative reliability
of the model (the model could easily be improved).  The physical
value of $\Lambda$ is determined in principle
by fitting to the derived
phenomenological parameters.  The
theory will contain the dynamical chiral symmetry breaking,
and will determine a chiral Lagrangian of the heavy--light system.

Upon Fierz rearrangement it is seen that
the interaction Lagrangian of eq.(2) contains
the  Nambu--Jona-Lasinio
model. The subsequent
analysis is standard. We can factorize eq.(2)
into a Yukawa theory  with a static
auxiliary field $\Sigma = \half(\sigma + i\pi^a\tau^a)$
on the scale $\mu\sim \Lambda$
and then  integrate out
the fermions to determine  the effective Lagrangian at scales $\mu < \Lambda$.
The field $\Sigma$ is $2\times 2$ complex at this
stage, which implies parity doubling of the $\pi$ and the parity
partner of $\sigma$, the $\eta$ is also present.
This analysis
is summarized in Appendix B.

The light sector effective Lagrangian at scales $\mu < \Lambda$
can be identified with a linear $\sigma$--model:
\bea
{\cal{L}}_L & = & \overline{\psi}(i\slash{\partial} - m_q )\psi
- \tilde{g}\overline{\psi}_L\Sigma_r\psi_R -
\tilde{g}\overline{\psi}_R\Sigma_r^\dagger\psi_L
\nonumber \\
& & + \Tr(\partial_\mu\Sigma_r^\dagger\partial^\mu\Sigma_r)
-  m_\sigma^2\Tr(\Sigma_r^\dagger\Sigma_r) +
\kappa\Tr (m_q\Sigma_r + h.c.)
\nonumber \\
& & +  \lambda \Tr(\Sigma_r^\dagger\Sigma_r
\Sigma_r^\dagger\Sigma_r)
\eea
$\Sigma_r$ describes the renormalized composite light mesons.
We have written the renormalized effective
Lagrangian, so that $\tilde{g}= g/\sqrt{Z_2}$.
${Z_2} = (g^2N/16\pi^2)\ln(\Lambda^2/\mu^2)$ is the finite,
induced  wave--function renormalization constant of the $\Sigma$ field.

  A $(\Tr\Sigma_r^\dagger \Sigma_r)^2$
term could be included in eq.(3), though it is subleading in $N_c$, and
for $SU(2)\times SU(2)$ with $(\sigma,\pi)$ real this
is equivalent to the quartic term we have included.
The theory can be tuned by choosing sufficiently
large coupling $g$ to develop a chiral symmetry breaking condensate,
thus generating a constituent quark mass.
The chiral symmetry breaking lifts
of the isovector, $0^+$ ($\Im(\pi)$) states The
$Re(\pi)$ $0^-$ pion, becomes the Nambu--Goldstone mode.
In QCD the residual $U(1)$ symmetry is broken by anomalies and the effects of
instantons.
This generates additional terms such as
an extra `t Hooft determinant, $\det\Sigma + h.c.$ term, which elevates
$\Im(\sigma)= \eta $.  Any additional necessary Wess--Zumino terms should be
incorporated as well.

Since the light sector dynamics is not our principal concern
in the subsequent analysis,
{\em we will henceforth assume that the fields $(\sigma, \pi^a)$ comprising
$\Sigma$
are real}, so $\Sigma$ henceforth contains only
the $0^-$ $\pi^a$ isotriplet and the real $\sigma$
isosinglet.   Therefore, eq.(3) becomes a linear version of the chiral
quark model ala Georgi, Manohar, and Holdom [14].
Nonetheless, we can dynamically put the
model either in a symmetric phase, $m_\sigma^2>0$, by choosing
${g^2N}/{4\pi^2}<1$, or
in a chiral symmetry breaking phase $m_\sigma^2<0$
with ${g^2N}/{4\pi^2}>1$. The critical bare coupling corresponds
to $m_\sigma^2=0 $ as $\mu_0\rightarrow 0$.
For further discussion of the light quark sector see Appendix B.

\vskip 0.1in
\noindent
{\em (ii) The Heavy--Light Quark Dynamics }
\vskip 0.1in

Now we focus on the dynamics of mesons containing one light
and one heavy quark.
The model produces one boundstate per channel in
the fermion bubble approximation.
We can conveniently solve the theory by
factorizing the heavy--light (HL) interaction into auxiliary
background interpolating fields with Yukawa couplings
to heavy and light quark vertices. The
original four--fermion interaction is recovered when the auxiliary
fields are integrated out.
Upon integrating out the quarks on scales  $\Lambda$
to $\mu $, the auxiliary fields acquire induced kinetic terms
on the scale $\mu$ and thus become
dynamical heavy--light mesons (``$B$--mesons'').  In this way
we  derive the effective Lagrangian for the HL mesons
coupled to the dynamical pions.

The  heavy--light fermion sector
interaction Lagrangian, together with the HQ kinetic
term, involves the HL cross--term of eq.(1)
and can be written as:
\bea
{\cal{L}}_{HL}& = & \overline{Q}(i\slash{\partial} - M) Q
-\frac{2g^2}{\Lambda^2}\overline{Q}
\gamma_\mu\frac{\lambda^A}{2}Q \; \overline{\psi}
\gamma^\mu\frac{\lambda^A}{2}\psi
\eea
Here we may generally take $Q=(t,b,c..)$ to be a multiplet of
$N_H$ heavy quarks, and $M$ the heavy quark mass matrix.  We
will presently consider, however, just a single heavy flavor in the following
discussion.
$g$ should be viewed as
the effective coupling at the scale $\Lambda$
in both the light sector and the HL sector of our model.
(In a more detailed discussion one might
wish to  distinguish the coupling
constant in the heavy--light effective action from that
of the light--light action at $\Lambda$; we ignore this
possibility in the present paper).

Upon Fierz--rearrangment of the interaction,
again keeping only leading terms in
$1/N_{C}$ and writing in terms of color singlet densities,
eq.(4) takes the form:
\bea
{\cal{L}}_{HL}& = & \overline{Q}(i\slash{\partial} - M) Q
+ \frac{g^2}{\Lambda^2}\left(
\overline{Q}^a\psi_{i}\overline{\psi}^iQ_a -
\overline{Q}^a\gamma^5\psi_{i}\overline{\psi}^i\gamma^5 Q_a \;\right.
\nonumber \\
& &
\left. -\frac{1}{2}\overline{Q}^a\gamma_\mu\psi_i\overline{\psi}^i\gamma^\mu
Q_a
-\frac{1}{2}\overline{Q}^a\gamma_\mu\gamma_5 \psi_i
\overline{\psi}^i\gamma_\mu\gamma_5 Q_a \right)
\eea
where $^i$ are the isospin indices, and $^a$ the heavy flavor
indices.

In the heavy quark limit we introduce a projection onto a heavy quark field
with a well defined four--velocity $v_\mu$.
Presently  we rewrite the full theory
identically in terms of a single four--velocity sector, corresponding
to the four velocity of the heavy constituent quark
or equivalently the boundstate heavy mesons:
\beq
Q \rightarrow \frac{1+\slash{v}}{2}\exp(-iMv\cdot x)Q(x)_v
\eeq
Note that $(1+\slash{v})Q_v/2 = Q_v$, i.e., the field $Q_v$
always carries an implicit factor of $(1+\slash{v})/2$.
The HQ kinetic term then takes the form:
\beq
\overline{Q}^a_v iv^\mu{\partial}_\mu Q_{va}
\eeq
The Isgur--Wise flavor symmetry is just the group of $SU(N_H)$
rotations on $Q^a_v$, and is now a manifest symmetry
of our Lagrangian.  We
will consider just a single heavy flavor in the following.

We now rewrite the terms of eq.(5) in a manifestly heavy
spin symmetric form, letting $Q\rightarrow Q_v$ and further
rearranging $\gamma$--matrices.  Then,
eq.(5) takes the form:
\bea
{\cal{L}}_{HL}& = & \overline{Q}_v iv^\mu{\partial}_\mu Q_{v}
+
\frac{g^2}{2\Lambda^2}\left(
\overline{Q}_v\psi_{i}\overline{\psi}^iQ_v -
\overline{Q}_v\gamma^5\psi_{i}\overline{\psi}^i\gamma^5 Q_v \; \right.
\nonumber \\
& &
\left. - \overline{Q}_v\gamma_\mu\frac{1-\slash{v}}{2}\psi_i
\overline{\psi}^i\frac{1-\slash{v}}{2}\gamma^\mu Q_v
+\overline{Q}_v\gamma_\mu \frac{1-\slash{v}}{2}\gamma_5 \psi_i
\overline{\psi}^i\gamma_5\frac{1-\slash{v}}{2}\gamma_\mu Q_v \right)
\eea
We have now brought the interaction to a form
which can be factorized by
introducing heavy static auxiliary fields, $(B,B')$.
To do so {\em we must
introduce four independent fields,} $B\; ( B^5)$ are $0^+$
($0^-$) scalars, while $B_\mu\; (B_\mu^5)$ are
$1^-$ ($1^+$) vectors. These form a minimal complete set
of auxiliary fields needed to factorize eq.(8) in the
HQ limit. Eq.(8) then becomes:
\bea
{\cal{L}}_{HL} & = & \overline{Q}_v iv^\mu{\partial}_\mu Q_{v}
+
{g}\overline{Q}_v\psi_{i}\overline{B}_v^i +
i{g}\overline{Q}_v\gamma^5\psi_{i}\overline{B}_v^{5i}
\nonumber \\
& &+{g}\overline{Q}_v\gamma_\mu \frac{1-\slash{v}}{2} \psi_i
\overline{B}_v^{i\mu}
- i{g}\overline{Q}_v\gamma_\mu\frac{1-\slash{v}}{2}  \gamma_5 \psi_i
\overline{B}_v^{5\mu} + h.c.
\nonumber \\
& &
- 2\Lambda^2 (\overline{B}^{i5}_{v}B_{vi}^{5} + \overline{B}_v^i{B}_{iv})
+ 2\Lambda^2 (\overline{B}_v^{i\mu 5}{B}_{iv\mu}^5 +
\overline{B}_v^{i\mu}{B}_{iv\mu})
\eea
Upon integrating out the $B$ fields in eq.(9) we reproduce
eq.(8).  (Note that the $B$ fields do not yet have canonical dimension
of heavy meson fields; see Appendix A).

Eq.(9) is a  heavy-spin symmetric form.  We can
assemble the auxiliary fields into
complex $\bf 4 $ multiplets under
$O(4)=SU(2)_h\times SU(2)_l$, where $SU(2)_h$ ($SU(2)_l$) is the
little group of rotations on $Q_v$ ($\psi$ and gluons)
which preserves $v_\mu$.
One heavy spin {\bf 4} multiplet
consists  of the $0^+$ scalar together with
the abnormal parity ($1^+$) vector as $(B, B^{5\mu})$
(the four--velocity label $_v$, and isospin $^i$
indices are understood):
\bea
{{B}}' & =  &(i\gamma^5 B + \gamma_\mu B^{5\mu})
\qquad
{\cal{B}}'  =  (i\gamma^5 B + \gamma_\mu
B^{5\mu})\left(\frac{1+\slash{v}}{2}\right)
\eea
The other $\bf 4 $ multiplet
consists of the usual $0^-$ scalar and a $1^-$ vector $(B^5, B^{\mu})$:
\bea
{{B}} & =  &(i\gamma^5 B^5 + \gamma_\mu B^{\mu})
\qquad
{\cal{B}}   =  (i\gamma^5 B^5 + \gamma_\mu
B^{\mu})\left(\frac{1+\slash{v}}{2}\right)
\eea
Under heavy spin $O(4)=SU(2)_h\times SU(2)_l$
rotations the $(B, B^{5\mu})$ mix analogously
to $(B^5, B^\mu)$. Note that $v_\mu B^\mu = 0$ always.
 We  have introduced the caligraphic ${\cal{B}}$ and
${\cal{B}}'$ with the explicit projection factors.
Falk has previously written similar effective ``superfields''
for excited mesons in model independent analyses; he
includes an extra factor of $\gamma^5$ (relative to us)
in his writing of effective
fields for the  $(0^+,1^+)$ multiplet in a model independent
approach [12]; for us the field ${\cal{B}}'$ has overall
odd parity while ${\cal{B}}$ is  even.

 The factorized heavy--light
interaction Lagrangian then takes the compact form:
\bea
{\cal{L}}_{HL}  & = &\overline{Q}_v iv^\mu{\partial}_\mu Q_{v}
+{g}\overline{Q}_{v}\;(-i\overline{\cal{B}'}^i  \gamma^5
+ \overline{\cal{B}}^i)\;\psi_{i} + h.c.
\nonumber \\
& & +
\Lambda^2 \left[\Tr({\cal\overline{B}B}) + \Tr({\cal\overline{B}'{B}'})
\right]
\eea
Notice that the combination $-i\gamma^5{\cal{B}'}^i
+ {\cal{B}}^i$ is coupled.
We emphasize that eq.(12)
is exactly equivalent to the full four--fermion theory
in the heavy quark symmetric and leading large--$N$ limit
eq.(4).
The theory forces a parity doubling of the heavy mesons upon
us because the chiral symmmetry is controlled dynamically
by $g$.  For weak $g$ the linear chiral
invariance is realized and the theory must contain
parity doubled meson states. Heavy spin symmetry
organizes the parity partners into heavy spin {\bf 4}--multiplets.
 The effect of chiral symmetry breaking
on the spectrum can now be investigated by solving the theory
and choosing the broken phase.

See Appendix A for a discussion of normalization conventions.

\vskip 0.2in
\noindent
{\bf III. Full Effective Lagrangian}
\vskip 0.05in

We now proceed to ``solve'' the theory.
The full effective Lagrangian for the heavy mesons is derived by
integrating out the heavy and light quarks in eq.(12) over momentum
scales $\mu < k < \Lambda$, keeping the leading
induced terms. Details of the
explicit calculations are given in  Appendix A.
 We begin the discussion with
the use of the linearly realized chiral symmetry
form, $\Sigma = \half(\sigma + i\pi\cdot \tau)$,
and we derive the nonlinear realization subsequently below. The
loop integrations result in an unrenormalized effective Lagrangian.  By
performing a conventional wave--function renormalization
and several field redefinitions we arrive
at the  full effective action valid to O$(\mu/\Lambda)^2$:
\bea
{\cal{L}}_{LH} & = & -i\half \Tr(\overline{{\cal{B}}}_{}v \cdot \partial
{\cal{B}})_{}
  -i\half \Tr(\overline{{\cal{B}}'}_{} v \cdot \partial
{\cal{B}}_{}')
\nonumber \\
& & -\frac{{g_{r}} }{4}
\left[  \Tr(\overline{{\cal{B}}_{}}\widetilde{\sigma}_{} {\cal{B}}_{})
 -\Tr(\overline{{\cal{B}}_{}'}\widetilde{\sigma}_{}{\cal{B}}_{}')
+
\Tr(\overline{{\cal{B}}_{}'} \pi_{}\cdot\tau  {\cal{B}}_{}) +
\Tr(\overline{{\cal{B}}_{}}  \pi_{}\cdot\tau {\cal{B}}_{}')\right]
\nonumber \\
& &
+\frac{h_{r}}{4\Lambda }
\left[ \Tr(\overline{{\cal{B}}_{}} (\widetilde{\sigma}_{}^2 + \pi_{}^2)
{\cal{B}}_{})
+ \Tr(\overline{{\cal{B}}_{}'} (\widetilde{\sigma}_{}^2  + \pi_{}^2)
{\cal{B}}_{}')\right]
\nonumber \\
&  &
+\frac{k_{r}}{4 f_\pi}
\Tr \left[
\overline{\cal{B}}\gamma^5
(\slash{\partial}\pi\cdot\tau ) {\cal{B}}
- \overline{\cal{B}'} \gamma^5
(\slash{\partial} \pi\cdot\tau) {\cal{B}'}
- \overline{\cal{B}'}\gamma^5
(\slash{\partial}\sigma)   {\cal{B}}
- \overline{\cal{B}}
\gamma^5(\slash{\partial}\sigma) {\cal{B}'}
\right]
\nonumber \\
& &
+\;\Delta_{} \left[\Tr(\overline{{\cal{B}}_{}}
{\cal{B}}_{}) + \Tr({\overline{{\cal{B}}_{}'}}
{{\cal{B}}_{}'})
\right]
\eea
The light quark PCAC masses are contained in
the  ``shifted'' $\sigma$ field,
$\tilde{\sigma} = \sigma + 2m_q\sqrt{Z_2}/g$.
The parameters of this Lagrangian are determined as:
\bea
g_r & = & \frac{g}{\sqrt{Z_2}};
\qquad
h_r  =  \frac{2g^2 \sqrt{Z_2}\Lambda}{Z_1} ;
\qquad
k_r  =  \frac{2g f_\pi \sqrt{Z_2}}{Z_1 }
\nonumber \\
\Delta & = & \frac{1}{Z_1}\left( \Lambda^2 - Z_1(\Lambda + \mu)/2\pi \right)
\eea
where:
\bea
Z_1 & = &
\frac{g^2N}{8\pi }( \Lambda  - \mu);
\qquad
 Z_2  =  \frac{g^2N}{16\pi^2}\left[\ln(\Lambda^2/\mu^2)\right]
\eea
The parameters defined above arise from the loop calculations
presented in Appendix A. The $g_r$, $h_r$ and $k_r$ are dimensionless.
They are determined
in principle by fitting the
observables of the model as in Section IV.  We will generally take
$\mu$ to be of order the light quark constituent mass, and it will
henceforth be neglected in the expression for $Z_1$.  Note that
terms like $\overline{\cal{B}'}\gamma^5
(\sigma\slash{\partial}\sigma) {\cal{B}}$ are potentially induced,
but they are subleading as $\sim O(1/\ln(\Lambda/\mu)$, relative to
the terms we keep.

We now identify the chiral representations of the composite fields
in the effective
theory. This can easily be done by returning
to eq.(12) and examining which heavy meson linear combinations
couple to $\psi_L$ and $\psi_R$.
If we define the following combinations:
\bea
& & {\cal{B}}_{1} = \frac{1}{\sqrt{2}}\left({\cal{B}} - i {\cal{B}}'\right)
\qquad
{\cal{B}}_{2} = \frac{1}{\sqrt{2}}\left({\cal{B}} + i {\cal{B}}'\right)
\eea
inspection of eq.(12) reveals that ${\cal{B}}_{1}$ (${\cal{B}}_{2}$)
couples to $\psi_R =(1+\gamma^5)\psi/2$ ($\psi_L=(1-\gamma^5)\psi/2$).
Thus, the chiral representation of ${\cal{B}}_{1}$ (${\cal{B}}_{2}$)
must be $(0,\half)$ ($(\half, 0)$).
Writing  in terms of $\Sigma = \half(\sigma + i\pi\cdot \tau)$
(we will henceforth ignore the $m_q$ contribution which
can easily be restored by shifting $\sigma\rightarrow \tilde{\sigma}
$), the effective Lagrangian becomes:
\bea
{\cal{L}}_{LH} & = & -i\half \Tr(\overline{{\cal{B}}}_1 v\cdot\partial
{\cal{B}}_{1} )
 -i\half \Tr(\overline{{\cal{B}}}_{2}  v\cdot\partial  {\cal{B}}_{2} )
\nonumber \\
& & -\frac{{g_{r}} }{2}
\left[  \Tr(\overline{{\cal{B}}_{1}}\Sigma^\dagger
 {\cal{B}}_{2}) +\Tr(\overline{{\cal{B}}_{2}}\Sigma
 {\cal{B}}_{1})
\right]
\nonumber \\
& & +\left(\Delta_{} + \frac{h_{r}}{\Lambda}\Sigma^\dagger \Sigma\right)
\left[  \Tr(\overline{{\cal{B}}_{1}} {\cal{B}}_{1})
+\Tr(\overline{{\cal{B}}_{2}} {\cal{B}}_{2})
\right]
\nonumber \\
&  &
+\frac{ik_r}{ 2f_\pi}
\left[  \Tr(\overline{{\cal{B}}_{1}}\gamma^5(\slash{\partial}\Sigma^\dagger)
 {\cal{B}}_{2}) - \Tr(\overline{{\cal{B}}_{2}}\gamma^5(\slash{\partial}\Sigma)
 {\cal{B}}_{1})
\right]
\eea
Inspection of the effective Lagrangian (as well as eq.(12))
confirms that it is manifestly  invariant
under $SU(2)\times SU(2)$ provided the fields transform as:
\bea
& & {\cal{B}}_{1} = (0, \half) \qquad {\cal{B}}_{2} = (\half,0)
\qquad \Sigma  = (\half,\half)
\eea
We now see that indeed, eqs.(13, 17) have a structure
analogous to that  of a parity doubled nucleon
theory, with ${\cal{B}}\sim (n,p)_{P=+1}$, the normal
even parity nucleon doublet, and
${\cal{B}}'\sim (n,p)_{P=-1}$ the odd parity doubling partner.
We give a brief synopsis of such a system at the end of
Appendix B.  The essential results are that the axial vector current
couples only through the perturbative $k_r$ term and describes
transitions between parity partners, and the parity degeneracy
will be lifted by $\VEV{\sigma}$.

Note that eqs.(13, 17) describe the heavy meson dynamics in
either a broken or an unbroken phase, \ie,  it is simply a linear
$\sigma$--model form.
In the spontaneously broken phase of the
heavy meson  theory we can pass to the
the nonlinear realization by replacing $\Sigma$ with
a unitary matrix field which is a function of angular
pion fields, and $\sigma$ is now decoupled.  Thus,
the nonlinear realization is:
\beq
\Sigma = \half f_\pi
\exp(i\pi\cdot\tau/f_\pi)  \qquad 2\Sigma/f_\pi =  \xi^2
\eeq
We can pass to the current form
by performing the chiral field redefinitions:
\bea
& & {\cal{B}}_{1} \rightarrow \xi^\dagger  {\cal{B}}_{1}\qquad
{\cal{B}}_{2} \rightarrow \xi  {\cal{B}}_{2}
\eea
We then have the Lagrangian:
\bea
{\cal{L}}_{LH} & = & -\half \Tr(\overline{{\cal{B}}}_{1}
v\cdot (i{\partial}
+ {{\cal{J}}}_L){\cal{B}}_{1} )
-\half \Tr(\overline{{\cal{B}}}_{2}v\cdot(i{\partial}+ {{\cal{J}}}_R)
{\cal{B}}_{2} )
\nonumber \\
& & -\frac{{g_r f_\pi} }{4}
\left[  \Tr(\overline{{\cal{B}}_{1}}
 {\cal{B}}_{2}) +\Tr(\overline{{\cal{B}}_{2}}
 {\cal{B}}_{1})  + h.c.
\right]
\nonumber \\
& & +\Delta_r
\left[  \Tr(\overline{{\cal{B}}_{1}} {\cal{B}}_{1})
+\Tr(\overline{{\cal{B}}_{2}} {\cal{B}}_{2})
\right]
\nonumber \\
&  &
-  i\frac{ k_r}{2}
\left[  \Tr(\overline{{\cal{B}}_{1}}\gamma^5\gamma_\mu {\cal{A}}^\mu
 {\cal{B}}_{2}) +\Tr(\overline{{\cal{B}}_{2}}\gamma^5\gamma_\mu {\cal{A}}^\mu
 {\cal{B}}_{1})
\right]
\eea
where:
\beq
\Delta_r = \left(\Delta_{} + \frac{h_{r}}{4\Lambda}f_\pi^2\right)
\eeq
and the chiral currents are:
\beq
{\cal{J}}_{\mu, L}  =  i \xi \partial_\mu \xi^\dagger \qquad
{\cal{J}}_{\mu,R}  = i \xi^\dagger  \partial_\mu \xi\qquad
{\cal{A}}_\mu  = \half({\cal{J}}_{\mu,R } - {\cal{J}}_{\mu,L} )\qquad
{\cal{V}}_\mu  = \half({\cal{J}}_{\mu,R }  + {\cal{J}}_{\mu,L} )
\eeq
As usual the ${\cal{J}}_A$ are matrices acting
on the isospin indices of meson fields.
The mass matrix of the chirally redefined
heavy mesons is at this stage non--diagonal.
We should mention that if an extra $\gamma^5$ were included
in the definition of the parity partner, then the axial current
components of the ${\cal{J}}_{\mu, L}$  and ${\cal{J}}_{\mu,R}$
terms would carry $\gamma^5$ factors, while no $\gamma^5$ would
occur in the $k_r$ term.

 Note that the
fields ${\cal{B}}_{1}$ and ${\cal{B}}_{2}$ are of mixed
parity.  The mass matrix
 can readily be diagonalized now that the Lagrangian is written
in the current form:
\beq
\widetilde{\cal{B}} = \frac{1}{\sqrt{2}} ({\cal{B}}_1 + {\cal{B}}_2)
\qquad
\widetilde{\cal{B}}' = \frac{i}{\sqrt{2}} ({\cal{B}}_1 - {\cal{B}}_2)
\eeq
with eigenvalues $2\Delta_r-gf_\pi/2$ and $2\Delta_r+gf_\pi/2$
respectively (recall that our normalization conventions
imply the physical mass shift
is $\delta M$ if
the Lagrangian contains $\half\delta M (\tr{\overline{\cal{B}}{\cal{B}}})$;
see Appendix A). The mass eigenfields are nontrivial
functionals of the pions through the absorbed $\xi$, $\xi^\dagger$ factors
as in eq.(20).
The Lagrangian now becomes:
\bea
{\cal{L}}_{LH} & = & -\half \Tr(\overline{\widetilde{\cal{B}}}
v\cdot (i{\partial}
+ {{\cal{V}}}){\widetilde{\cal{B}}} )
 -\half \Tr(\overline{\widetilde{\cal{B}}}'
v\cdot (i{\partial}
+ {{\cal{V}}}){\widetilde{\cal{B}}}' )
\nonumber \\
& & +\left( \Delta_r-\frac{g_r f_\pi}{4}\right)
\Tr\overline{\widetilde{\cal{B}}}
 \widetilde{\cal{B}}
+\left( \Delta_r+\frac{g_r f_\pi}{4}\right)
\Tr\overline{\widetilde{\cal{B}}}'
 \widetilde{\cal{B}}'
\nonumber \\
& & -\half \Tr(\overline{\widetilde{\cal{B}}}
(v\cdot {{\cal{A}}}){\widetilde{\cal{B}}}' )
 -\half \Tr(\overline{\widetilde{\cal{B}}}'
(v\cdot {{\cal{A}}}){\widetilde{\cal{B}}} )
\nonumber \\
& & - i\frac{k_r}{2}
\Tr \overline{ \widetilde{\cal{B}}}\gamma^5 \slash{{\cal{A}}}
\widetilde{\cal{ B}}
+ i\frac{k_r}{2}
\Tr \overline{ \widetilde{\cal{B}}}'\gamma^5 \slash{{\cal{A}}}
\widetilde{\cal{ B}}'
\eea
Note the appearance of the off--diagonal pionic transition terms of the form
$\overline{\widetilde{\cal{B}}}
(v\cdot {{\cal{A}}}){\widetilde{\cal{B}}}'$.
\footnote{The coefficient of this term corresponds to $h=1$
and  $k_r = g$ in
Falk's notation [12]. Our conclusion is that $k_r=g_A < h=1$,
and following Falk's analysis the
chiral perturbative contribution of these resonances, \eg,
to $f_{Ds}/f_{D}$, is significant. }
At this stage it can be seen that these terms are associated with
a Goldberger--Treiman relation, by taking
${\cal{A}}_\mu = \partial_\mu \pi\cdot \tau/f_\pi$, integrating
by parts, and using the equations of motion.  One finds
that  the ${\cal{B}}'{\cal{B}}\pi$
amplitude has a coupling strength $g_{B B'\pi} = g_r$, and this is
seen to be  given by
$\Delta M/f_\pi$.

 We can decouple the heavier field $\widetilde{\cal{B}}'$ to leading order
in the mass gap $g_r f_\pi$ by ``integrating it out''
(which amounts to setting it to zero in leading order).
We can then perform the residual mass redefinition:
${\cal{B}}\rightarrow \exp(- iM v\cdot x){\cal{B}}$ where
$M = 2\Delta_r- {g_r f_\pi}/{2}$
to yield the final result:
\bea
{\cal{L}}_{LH} & = & -\half \Tr \overline{\widetilde{\cal{B}}}i
v \cdot({\partial}
+ {\cal{V}}){\widetilde{\cal{B}}}
-i\frac{g_A}{2}
\Tr \overline{ \widetilde{\cal{B}}} \gamma^5 \slash{{\cal{A}}}
\widetilde{\cal{ B}}
\eea
where we now discover that:
\beq
g_A = k_r = 2g f_\pi \sqrt{Z_2}/Z_1
\eeq
Our fit to the model yields $g_A=0.32$ (see
eq.(38) and discussion).
Eq.(27) is equivalent to the point of departure taken by
ref.[6] in writing effective Lagrangians
involving simultaneous heavy symmetries and chiral symmetries.
Use of this effective Lagragian
is justified so long as $q$ is small
compared to the mass gap.   We see that $g_A$ here arises from the
perturbative $k_r$ term, which is subleading to unity in our
expansion.

In summary, the central observation of this analysis is that the
underlying chiral representations
of the full HL meson theory is a parity doubled scheme.
There are two general implications of such a scheme:
(1) The mass gap between the parity partners
arises from $VEV{\sigma}$.  Thus a Goldberger--Treiman relation refers,
not to the overall mass of the $B$ mesons $\sim M$,
but rather to the mass
splitting between the even and odd parity multiplets:
\beq
\Delta M =   g_{BB'\pi} f_\pi
\eeq
Here $g_{BB'\pi}=g_r$ is the $BB'\pi$
transition coupling constant and is the analogue of
the $g_{NN\pi}$ in the nucleon system.  We note that the light
quark constituent mass is given by $m_c \approx  g_r f_\pi/2$
so we expect $\Delta M \approx 600$ MeV, however
this must be obtained in principle
from a fit of the model to all data
(see section  IV.(ii);
Unfortunately, without exceptional circumstances
the width of this state is too large for observation.) (2) $g_A$ is
{\em not necessarily} expected to be $\sim 1$, being given by
a subleading perturbative contribution, $k_r$, alone.  This is
essentially a consequence of parity doubling and contrasts the
or the chiral quark model in which, $g^q_A= 1$ is a leading term.
 The fit we present
below in section IV(ii),  which is crude, yields $g_A \approx 0.32$.
This result may be indicated in the $D$--system where
$D^*\rightarrow D + \pi$ gives $g_A < 0.7$ [6, 11].

\vskip .1in
\noindent
{\bf IV. Other Observables: $f_B$, and Isgur-Wise function}
\vskip .05in
\noindent
{\em (i) Heavy Meson Decay Constant, $f_B$}
\vskip .05in
\noindent

We presently compute the heavy meson decay constant $f_B$.
Consider the heavy--light
axial current  $\overline{Q}\gamma_\mu\gamma^5\psi $.  We
can compute the renormalized matrix element:
\beq
\sqrt{Z_1}^{-1}\int d^4x\; e^{-iMv\cdot
x}\bra{{\cal{B}}}\overline{Q}(x)\gamma_\mu\gamma^5\psi(x)
\ket{0} = f_B \sqrt{M_B} v_\mu   + ...
\eeq
As a consequence of the heavy quark spin symmetry,
$B^5$ and $B_\mu^5$
have identical decay constants for the axial vector current,
while $B$ and $B_\mu$ have the same decay constants
for the vector current.
The $B$--meson must have a properly normalized
kinetic term, which includes the finite renormalization effects,
${\cal{B}}\rightarrow \sqrt{Z_1^{-1}}{\cal{B}}$.
We adopt a conventional normalization in which
we expect $f_B \approx 180$ MeV.

The amplitude on the {\em lhs }
takes the form:
\bea
&  & \frac{igN}{2\sqrt{Z_1}}\int \frac{d^4 k}{(2\pi)^4}
\Tr[\gamma_\mu \gamma^5 (\slash{k}-\slash{p} + m_q) (1-\slash{v})
(i\gamma^5B^5 + \gamma_\nu B^\nu)]\frac{1}{((k-p)^2 - \Omega^2)}
 \frac{1}{(v\cdot k) }
\nonumber \\
& = & \frac{2gN}{16\pi^2\sqrt{Z_1}}
v_\mu
{{B}}^5 \left[ \Lambda^2 - \mu^2
+ \pi v\cdot p (\Lambda - \mu) + \half \pi \widetilde{\sigma} \Lambda
- \Omega^2\ln(\Lambda^2/\mu^2)+ O((v\cdot p)^2) \right]
\nonumber \\
\eea
We see that the integral involved here is identical to
$I_1$ of eqs.(60, 62), and thus the equations of motion can be used
for the $B^5$ fields.
Upon  use of the equation
of motion, shifting $v\cdot p \rightarrow  2\Delta + ...
= 2\Lambda^2/Z_1 - \Lambda/\pi + ...$  a large cancellation
is seen to occur on the {\em rhs} of eq.(30) leaving:
\bea
&\rightarrow  &  \frac{2gN}{16\pi^2\sqrt{Z_1}}
v_\mu
 \left[ 2\pi\Lambda^3/Z_1
\right]\;
= \;   \frac{2}{g}
v_\mu
 \left[ \Lambda^2/\sqrt{Z_1}
\right]
\eea
We thus obtain:
\beq
\sqrt{M_B}f_B = \left(\Lambda \right)^{3/2}\frac{4\sqrt{2\pi}}{g^2\sqrt{N }}
\eeq
For example, let $g^2N/4\pi^2 =1$ and
use $f_B=180$ MeV, $M_B = 5$ GeV as input parameters, to find
$\Lambda= 1.35$ GeV for $N_C=3$.
Remarkably, our result is insensitive to the light
quark masses.

$f_B$ is a measure of the wave--function of the meson at
the origin in a nonrelativistic potential model.  We can
compute the wave--function in
principle in our model by point--splitting the current
in eq.(29):
\beq
\sqrt{Z_1}^{-1}\int d^4x\; e^{-iMv\cdot
x}\bra{{\cal{B}}}\overline{Q}(x-\epsilon/2)\slash{v}\gamma^5\psi(x
+\epsilon/2)
\ket{0} = {\cal{N}}\Psi(\epsilon)
\eeq
where ${\cal{N}}$ is a normalization factor.
This essentially replaces the momentum space cut--off procedure
by a point--split regulator, and $\Lambda \sim 1/\epsilon$.
The wave--function is
singular at the origin, and is not normalizable without a
spatial cut--off of the normalization integral at $1/\Lambda$
(our theory makes no sense at shorter distances than this).
Thus the wave--function is at the origin is given  effectively by:
\beq
|\Psi(0)| \sim \Lambda^2/\sqrt{Z_1}\sim (\Lambda)^{3/2}.
\eeq
This implies that the
result for $f_B$ is insensitive to infrared parameters
such as the light quark masses in our model, and indeed we find $f_{B,s}
= f_{B,u}$.  This is a defect of the model, but it
is an expected result of an extremely relativistic, potential dominated system.
 In this sense, QCD lies
somewhere between this extreme result and that of a naive potential model.

\vskip .1in
\noindent
{\em (ii) Fitting the model to data:}
\vskip .1in
\noindent

While  the model we have presented is not likely
to be quantitatively successful, we can  attempt
a fit to  observables, and  predict some features
of the HL meson system.
 We use as independent inputs
$f_\pi=95$ MeV, and  $f_B = 180$ MeV for
$M_B=5$ GeV. The latter
implies $\Lambda = 1.35$ GeV as discussed in
the previous subsection.
We see, owing to the smallness of
the ratio $f_\pi^2/\Lambda^2= (\kappa - 1)/g^2$,
that  $\kappa = g^2N/4\pi^2$
is very close to unity. In defining $Z_2$ we cut--off the renormalization group
flow at an  infrared scale $\mu\sim m_c$
taken as the approximate constituent
light quark mass.
Then, to obtain $Z_2 = \frac{1}{4}\kappa\ln{\Lambda^2/m_c^2} $ we
self--consistently
solve for the constituent quark mass $m_c = \half g f_\pi/ \sqrt{Z_2(m_c)}$.
This yields:
\beq
\frac{g^2 N}{4\pi^2} = 1.065;\qquad
g= 3.75\qquad
\Lambda = 1.35 \;\makebox{GeV}\qquad
Z_2 = 1.11\qquad
m_c = 169\; MeV
\eeq
$m_c$ is about a factor of two too small.
We can moreover use
the pion mass, $m_\pi$,
 to extract the light quark PCAC masses:
\beq
m_\pi^2 f_\pi^2 = \omega (m_u + m_d) f_\pi
\qquad
\omega = \frac{gN\Lambda^2}{8 \pi^2 \sqrt{Z_2 }} = 0.25 \;(GeV)^2
\eeq
hence, $m_u + m_d = 8.6$ {MeV},
which is to be compared with the conventional $\sim 15$
MeV, and is small.  Also, $m_s \approx m_K^2(m_u + m_d)/m_\pi^2
\approx 107$ MeV is small.

The mass gap between the excited $0^+$ and groundstate
$0^-$ mesons is
then:
\beq
\Delta M = g_r f_\pi \approx 2m_c \approx 338\;\makebox{MeV}
\;\;[600\;\makebox{MeV}]
\eeq
The result in brackets obtains when the known constituent masses
are inputted.  The decay width $\Gamma(0^+\rightarrow 0^-1)$
is given by $(\Delta M /f_\pi )^2 |k_\pi|/8\pi$. This is much
too large for observation of these
resonances when $k_\pi \sim \Delta M \sim 600$ MeV; with the lower
estimate of $\Delta M \sim 338$ MeV the width approaches $150$ MeV,
which is still too large.  Hence, the direct observation of the
parity partners of the groundstate is unlikely. Their effect in chiral
perturbation theory is nontrivial [12]; conceiveably the decay width
$\Gamma(D_s(1^+)\rightarrow D_{u,d}^*(1^-) + K)
\sim |p_K|$ is kinematically
suppressed by the $K$--meson mass and the $1/M$ corrections
to the $D$ masses.

We obtain the axial coupling constant:
\beq
g_A = \frac{2g f_\pi \sqrt{Z_2}}{Z_1 } =
\frac{4f_\pi \sqrt{\ln(\Lambda^2/m_c^2) } }{\Lambda \sqrt{N }}
\approx 0.32
\eeq
We might expect both $\Delta M$ and $g_A$ to be  underestimated
in this approximation, as are the light sector observables,
owing largely to the short--distance singularity of
our wave--function.

$g_A$ can be
in principle extracted from the decay $D^*{}^+\rightarrow D^0 + \pi^+$,
though it is unmeasured to date.
This decay partial width is given by ref.[6], and
in our conventions it takes the form:
\beq
\Gamma = \frac{g_A^2}{12\pi f_\pi^2}|p_\pi|^3
\eeq
where $p_\pi \sim 38.9$ Mev.  While this
width is not yet measured directly, we can
use the analysis of  ref.[19] to obtain an estimated result
 of $\Gamma = 53.4$ KeV  from the
measured branching ratio of $(D^*\rightarrow D\gamma)/(D^*\rightarrow D\pi)$
and a potential model calculation of $D^*\rightarrow D\gamma  $.
Combining these results we find  $ g_A =  0.56$,
which is compatible with the parity doubled interpretation,
but is also not far from the naive $g_A\sim g^q_A \sim 0.8$ from the chiral
quark model (note that we derive the chiral light--quark model here
with $g^q_A=1$, thus our prediction of $g_A\sim 0.3$
represents a significant suppression).
Amundson et.al. [11] give the current experimental limit of $g_A < 0.7$
consistent
with this result.  Thus, our model indicates that
$g_A$ is suppressed and smaller than unity, giving the physical
underlying rationale, though the situation is arguably not
decisive.

Note that $Z_1=\pi \Lambda/2 \sim 2.12$ GeV and $Z_2
\sim 1.1$.
Hence, $2\Delta = 3\Lambda/\pi \approx 1.3$ GeV.
Our model seems to
suffer from generating a value of $\Lambda$ that is slightly
large. This implies
$\epsilon = 4Z_2\Delta/Z_1\approx 1.2$,  suggesting
that our approximation of truncating on the $Z_2 (v\cdot p)^2/Z_1$
terms is probably unreliable (Appendix A).

The binding energy is determined in the model. Neglecting the
light quark PCAC masses we have in the infinite mass HQ limit:
\beq
M_{D,B} = M_{c,b} + \delta m;\qquad \delta m =
 2\Delta - m_c + \frac{2h_r m_c^2}{g_r^2\Lambda}
\eeq
where $m_c$ is the constituent light quark mass (the latter
term is small, but non--negligible).
 For the fit we have presented we find $h_r = 2g^2\sqrt{Z_2}\Lambda/Z_1
= 17.9$, and $g_r = g/\sqrt{Z_2} = 3.55$.  If we use
a conventional charm (b--) quark mass of $1.2-1.8$ GeV, ($4.5-5.0$ GeV)
this overestimates: $M_D \approx 2.4-3.0$ GeV, ( $M_B = 5.8-6.3$ GeV).
These results should be corrected for finite mass of the heavy
quark.  The corrected boundstate mass is:
\beq
M_B =  \sqrt{M_Q^2 + 2M_Q\delta m}
\eeq
This yields a result $M_D \approx 2.0-2.75$ GeV, ( $M_B = 5.65-6.3$ GeV).
This illustrates the problem of $\Lambda$ being too large in the
model.

The effect of the explicit $SU(3)$ breaking light quark masses
is calculable, upon restoring these terms in eq.(13) as
contained in the shifted $\tilde{\sigma}$ field.
Using the full constituent quark mass $m_c = g_r\tilde{\sigma}/2$
we have:
\bea
M_{Bq} - M_{Bq'} & = & -(m_c- m_{c'})+ (m_c^2- m_{c'}^2)\left(\frac{2 h_r
}{g_r^2\Lambda}\right)
\nonumber \\
& = & -(m_c- m_{c'})+ (2.1\times 10^{-3}\;(MeV)^{-1})(m_c^2- m_{c'}^2)
\eea
For the $D_s$--$D_0$ mass difference we take $m_c = 450$ MeV
(the strange quark constituent mass) and $m_{c'} = 300$ MeV
to obtain $M_{Ds}-M_{D(u,d)} = 86.25$ MeV. (If we use the predicted
$m_{c'} = 169$ MeV and $m_c = 276$ MeV we obtain $52.5$ MeV).
This compares to $\approx 100$ MeV experimentally.  It shows,
however, that the model must include the effects of
the $\sigma^2$ term in computing these differences.
The $M_{D+}-M_{D0} = [(+0.26),\; (-0.3)](m_d-m_u) \approx (2.6,\; -3)$ MeV
(using standard constituent masses in the first entires, and
the model's derived constituent masses in the second).
This is subject to
electromagnetic corrections, estimated  to  be $+ 2.0 $ MeV.

We have seen that $f_B$ is insensitive to the light quark masses
in this model.
Thus, we obtain
$f_{Bs}/f_{Bu,d}= 1 $, while lattice
results yield $\sim 1.09$ [20].
This result owes to the
unrealistic non--normalizeable singularity of the wave--function at the
origin.
This is consistent with the behavior of the binding energy
for small constituent quark mass, in which increasing the constituent mass
actually decreases the meson mass
(for large constituent mass the $\sigma^2$ terms
contribute to increase the meson mass).

\vskip .1in
\noindent
{\em (iii) Isgur--Wise Function}
\vskip .1in
\noindent

The analysis of the Isgur--Wise function in the model involves
a careful treatment of the cut--off procedure. We select
a preferred cut--off by demanding the validity of reparametrization
invariance (or the residual mass symmetry) [15-17].

We consider the transition amplitude in 4--velocity defined by
the matrix element  $\bra{B_v}\overline{Q}_v\Gamma Q_{v'}\ket{B_{v'}}$,
where $\Gamma$ is an arbitrary Dirac matrix.
\bea
I & = & \frac{\tilde{g}^2}{M} \int \frac{d^4 k}{(2\pi)^4}
\Tr\left( \overline{B} \frac{1+\slash{v}}{2}
 \frac{\slash{k}}{(k^2 - m^2)}
\frac{1+\slash{v}'}{2}
B \Gamma \right) \frac{1}{v\cdot k } \frac{1}{v'\cdot k }
\eea
This involves the integral:
\bea
I_1 & = & \int \frac{d^4 k}{(2\pi)^4}
\frac{{k}_\mu}{(k^2 - m^2)}
 \frac{1}{v\cdot k } \frac{1}{v'\cdot k }
\nonumber \\
& = & A\cdot (v^\mu + v'{}^\mu)
\eea
where the latter term follows
by symmetry, since $A$ can only depend upon $v\cdot v'$ and is
even under $v \leftrightarrow v'$.  Now multiply by
$v + v'$:
\bea
2A\cdot(1+v\cdot v') & = & \int \frac{d^4 k}{(2\pi)^4}
\frac{k\cdot(v + v')}{(k^2 - m^2)}
 \frac{1}{v\cdot k } \frac{1}{v'\cdot k }
\nonumber \\
& = & \int \frac{d^4 k}{(2\pi)^4}
\frac{2}{(k^2 - m^2)}
 \frac{1}{v\cdot k }
\nonumber \\
& = & \frac{i}{8\pi}\left( \Lambda - 2m\right)
\eea
Therefore:
\bea
A & = & \frac{i}{16\pi(1+v\cdot v')}\left( \Lambda - 2m\right)
\eea
and we conclude that the Isgur--Wise function is
given by:
\beq
\xi(v\cdot v') = \frac{2}{1 + v\cdot v'}
\eeq
This should be true if the
momentum space integral is Lorentz--invariant and finite.
Computing the integral directly, without recourse
to the symmetry argument one can obtain:
\beq
\frac{1}{2}\xi(v\cdot v') =  \int_0^{\pi/2} d\theta
\frac{\cos\theta}{(
1+2v\cdot v'\cos\theta\sin\theta )^{3/2} }
\eeq
which agrees with the previous result.

This result contains a t--channel pole
at $t=-(M_1+M_2)^2$, where $M_1$ ($M_2$) is the
incoming (outgoing) heavy meson mass. One might
ask if this is consistent with the slope constraint of
de Rafael and Taron  [18]
arising from $t$--channel unitarity?
Our slope, $\xi'(0) = -1$ is inconsistent with the
their lower bound of $-1/2$ arising from a $t$--channel branch--cut
at threshold. Grinstein and Mende [18] have pointed out
that the de Raphael--Taron constraint is weakened
by effects of resonance poles, as we are presently observing.
However, the $t$--channel unitarity constraint is an
interesting issue in
HQET.  In an HQET such as we have studied, the anti--particle has
been discarded at the outset, and with it goes crossing symmetry
and $t$--channel unitarity. Moreover, our cut--off theory
would seem to require the bound
of $Q^2 < \Lambda^2$ without a unitarization.  Since
$Q^2 = 2M^2(1-v\cdot v')$, we see that this bound corresponds
to the limit $v\cdot v'\rightarrow 1$ for $M\rightarrow \infty$.
Nevertheless we can compute the $t$--channel behavior by incorporating
the heavy anti--quarks  and computing the large--$N$ bubble sum
with the full interaction. While we do not
present this analysis here, we find, perhaps not surprisingly,
a Nambu--Jona-Lasinio pole at $M_1 + M_2$ is generated, and
our slope is consistent with the existence of this pole.

The previous result of eq.(47) is, however, sensitive to the definition of
the cut--off procedure, which we have taken to be
a Lorentz--invariant Euclidean momentum space cut--off.
Different results follow if the energy integrals are first
performed by residues, and then a 3--momentum cut--off
procedure is used.
To see this let us compute directly with a 3--momentum cut--off
Let $v' = (1,\vec{0})$ and $v=(v_0, \vec{v})$.
First we perform the energy integral
by closing $dk_0$ below to pick
up the single pole:
\bea
A(1+v\cdot v') & = & \int \frac{d^4 k}{(2\pi)^4}\left(
\frac{k_0 + v_0k_0-\vec{v}\cdot\vec{k}}{k_0^2 - \vec{k}^2 + i\epsilon}\right)
 \left(\frac{1}{v_0 k_0 - \vec{k}\cdot\vec{v} }
 \right)\left( \frac{1}{k_0 }\right)
\nonumber \\
& = &
\frac{i}{4\pi^2}
 \int_0^\infty d k\; \left[\frac{1}{\sqrt{(v\cdot v')^2 - 1}}
 \ln\left(\frac{ v\cdot v'+ \sqrt{(v\cdot v')^2 - 1}}{ v\cdot v' -
\sqrt{(v\cdot v')^2 - 1} }\right)+ 2 \right]
\nonumber \\
\eea
This result, using a non--Lorentz
invariant regularization
procedure, differs significantly from eq.(47) in which the Lorentz
invariant cut--off was used.

There is however an implicit gauge invariance in heavy
quark effective theories associated with the ``residual mass
ambiguity.''  One is free to add a term $\chi\overline{Q}_vQ_v$
to the effective Lagrangian of eq.(12).  $\chi$ should be viewed
as a gauge potential in the sense that if we redefine the
heavy quark mass $M\rightarrow M +\mu$, and thus $Q\rightarrow
\exp(i\mu v\cdot x)Q$ we can compensate this gauge transformation
by shifting $\chi \rightarrow  \chi + \mu$.  Hence
$iv\cdot \partial + \chi$ is a covariant derivative.  This
is essentially the demand that the global zero of energy of
a classical theory be arbitrary.  This symmetry and its implications
will be discussed elsewhere, however we can see immediate
implications for our present problem.

We can observe that the non--Lorentz
invariant regularization  procedure violates the $\chi$ symmetry.
Consider the integral involved in our calculation
of the Isgur--Wise function:
\beq
I = \int\frac{d^4k}{(2\pi)^4} \frac{1}{[(k+p)^2-m^2](v\cdot k + \mu)}
\eeq
We have chosen to route the external momentum
$p$ through the light fermion line.
The $\chi$ symmetry applies to the external
heavy mesons and requires that the
following shift in $v\cdot p$ and
$\mu$ be a symmetry of the integral:
\beq
v\cdot p \rightarrow v\cdot p + \chi;
\qquad
\mu \rightarrow \mu + \chi
\eeq
This is readily seen to be a symmetry in the case of
the momentum $p$  routed
through the heavy fermion line.

In the present example we can implement this by shifting $p \rightarrow p +
v\chi$.
Therefore the shift in the integral is:
\beq
\frac{\delta}{\delta\chi}I
= \int\frac{d^4k}{(2\pi)^4}
\left( \frac{-2v\cdot k}{[(k+p)^2-m^2]^2(v\cdot k + \mu)}
-\frac{1}{[(k+p)^2-m^2](v\cdot k + \mu)^2}
\right)
\eeq
The symmetry condition is $\frac{\delta}{\delta\chi}I = 0$,
and is equivalent to demanding that
the integral generates no nontrivial surface term upon
shifting $k\rightarrow k + a$.
For simplicity we consider the surface term:
\beq
S
= \int\frac{d^4k}{(2\pi)^4}
\left( \frac{-2v\cdot k}{(k^2-m^2)^2(v\cdot k )}
-\frac{1}{(k^2-m^2)(v\cdot k )^2}
\right)
\eeq
If we evaluate $S$ using the covariant cut-off we find
that $S = 0$.

Now consider computing $S$ by first performing $dk_0$ by
residues, then the residual 3--momentum integration with a
cut--off.  We find:
\bea
\int\frac{d^4k}{(2\pi)^4}
 \frac{v\cdot k}{(k^2-m^2)^2(v\cdot k )}
& = &
\int\frac{d^4k}{(2\pi)^4}
\frac{1}{(k^2-m^2)^2}
\nonumber \\
& = &
\frac{i}{4\pi^2} \int_0^\infty
\frac{k^2 dk}{({k}^2+m^2)^{3/2}}
\eea
Consider now
\bea
\int\frac{d^4k}{(2\pi)^4}
 \frac{1}{(k^2-m^2)(v\cdot k )^2}
& = &
\frac{-i}{8\pi^2} \int_0^\pi \sin\theta d\theta \int_0^\infty
\frac{k^2 dk}{({k}^2+m^2)^{3/2}(v_0 - |\vec{v}|u(k)\cos\theta) }
\nonumber \\
\eea
where $u(k) = k/\sqrt{k^2+ m^2}$ and $v_\mu = (v_0, \vec{v})$,
and thus $v_0^2 - \vec{v}^2= 1$.
Notice that if either $m\rightarrow 0$ or if
$\vec{v}\rightarrow 0$ then $S \rightarrow 0$.  Let us expand in $m^2$,
using the latter results, to find for $S$:
\bea
S &= & \frac{i}{4\pi^2}(1- v_0^2)\int_0^\infty\frac{x^2 dx}{(1+ x^2)^{5/2}}
\nonumber \\
 &= & \frac{i}{12\pi^2}(1- (v\cdot\eta)^2)
\eea
Here we introduce a four--vector $\eta_\mu = (1, \vec{0})$ which
is the direction of the $dk_0$ line integration.

This latter result implies that the $\chi$ symmetry is broken when the
$k_0$ line integral is not parallel to $v_\mu$.  For the
computation of the Isgur--Wise function where $v\neq v'$ then
the $\chi$ symmetry can never be present in the
residue computation.  However, utilizing the Lorentz invariant
cut--off we see that the $\chi$ symmetry can be maintained.
The $\chi$ symmetry therefore requires
that we reject the result of eq.(49) in favor of
eq.(47) which is consistent with the absence of momentum
space surface terms, and the attendant symmetry.

\vskip .1in
\noindent
{\bf VI. Conclusions}
\vskip .1in

We have presented perhaps
the simplest, solvable, strongly coupled toy
field--theoretic model in which both
heavy quark and chiral symmetries
are present at the fundamental quark level, and the dynamics of chiral
symmetry breaking is made explicit.
We find that the chiral representations of the heavy mesons are
parity doubled.  This has a well defined meaning in
the toy model because we can tune the coupling constant to restore
the spontaneously broken chiral symmetry.  In the symmetry limit
the groundstate is a degenerate system of $(0^-, 1^-)$ and
$ (0^+, 1^+)$ heavy mesons.  When chiral symmetry is broken the degeneracy
is lifted, elevating the $ (0^+, 1^+)$ and depressing the $(0^-, 1^-)$
heavy meson multiplets.
  We obtain the full chiral Lagrangian containing the
parity doubled composite HL mesons together with the composite
pions.
 The mass gap between the multiplets is given by $g f_\pi$,
and the analogue of the Goldberger--Treiman relation of the system reflects
this, $g_{B B'\pi} = \Delta M/ f_\pi$.  We are able in
the broken phase to pass to a nonlinearly realized chiral symmetry,
and to write a purely derivatively coupled pion effective
Lagrangian.  We can then decouple the heavier parity doubling field to arrive
at the conventional low energy effective chiral Lagrangian
for ${\cal{B}}$.

We believe that the general phenomenon of the
parity doubled chiral representations of heavy mesons
is inherent to QCD itself.
We emphasize at the outset
that this toy model is intended to convey the schematics of
QCD chiral dynamics in heavy--light mesons
 The model is designed to
imitate these dynamical features of QCD, rather than
provide a detailed phenomenological fit.
Nonetheless, the simplest fit seems to agree
within a factor of two to the expected values of physical
quantities, and is predictive.  While we would be inclined
to trust the result $g_A =0.32$ only to within a factor of $2$, the
model suggests that $g_A$ is smaller than might be naively expected
on the basis of the simple constituent quark model in which.
The direct observation of the parity partners of the groundstate mesons
is unlikely owing to their large widths.  It would be interesting
to extend these results to the heavy quark containing
baryons where similar conclusions must hold.

Our analysis achieves the basic systematics of chiral symmetry
in these systems where we might expect potential models
to fail.  The chiral symmetry limit is
relativistic, and the chiral symmetry breaking is a dynamical
rearrangement of the vacuum, two features which would be hard to realize
in any potential model treatment.
One must be careful in estimating the value of $g_A$ in a naive potential model
unless the mixing with the parity doubled states is under control. As we
have observed in eq.(21), the  $g_A$ term is a transition matrix element
between the $0^+$ and $0^-$ states in the mixed parity
basis appearing there.  In a basis
in which the $g_A$ term is diagonal, the mass matrix must be correspondingly
diagonal. There remains the transition amplitude term between the parity
partners (some authors include an extra factor of $\gamma^5$ in
the odd parity fields, and this transition term can then be mistaken for
the $g_A$ term in a mixed parity basis).
 In our model, the constituent quarks are found to have
$g^q_A=1$, and yet the value of $g_A$ obtained in the Lagrangian
of ref.[6] is
suppressed to $\sim 0.32$. This is a subtlety of parity doubling which
must be treated with some care. The resonances may have important
contributions in chiral perturbation theory to quantities such
as $f_{Ds}/f_D$ and flavor ratios of Isgur--Wise functions [12]
(in the notation of Falk, $h=1$ and $g_A=g$, and thus $h^2>>g^2$
in our model, so the resonance contributions are significant).

We have studied the physical predictions of this system.
The wave--function of the theory is too singular at the origin
to represent a realistic QCD wave--function.  This is a consequence of
the strong coupling of the point--like
four--fermion interaction term. While
it is a defect of the model, it indicates the trend in a theory in which
the potential term is dominating the dynamics.  For example, we obtain the
unrealistic
$f_{Bs}/f_{Bu,d} = 1$, because the singular short--distance behavior
of the wave--function becomes insensitive to the infra--red parameters
of the theory. This contrasts lattice results, indicating $f_{Bs}/f_{Bu,d} =
1.1$ [20]. However,  a weakly coupled potential model would give the
larger result
$\sqrt{m_s/m_d} \sim 1.2$ [21].

In our analysis we
fix $\Lambda \sim 1.35$ GeV from $f_B$ and examine the relationship
with the cut--off wave--function at the origin. Inputting also $f_\pi$ fixes
$g$, and marginal results
(within a factor of 2) obtain for $\Delta M$, $g_A$ and the light quark
sector. A  defect, related to the short--distance
singularity of the wave--function, is the fact that for small
light quark constituent mass, the groundstate mass is actually depressed
as the  light quark constituent mass is increased from small
constituent mass.
Nonetheless, the common $h_r$ term is sufficiently large for
$m_c\sim 300$ MeV that a reasonable result for $M_{Bs}-M_{Bu,d}$ emerges
from the fit.

Of further interest is the Isgur--Wise function, which is
associated with an ambiguous linearly divergent integral
in the present scheme. The ambiguity is resolved by invoking
``residual mass invariance'' [15,16],
or equivalently, ``reparameterization invariance,''
and enforcing an associated
Ward identity [17].  The simple Isgur--Wise function corresponds
to a $t$--channel threshold pole at $(M_1+M_2)^2$.  This pole is beyond
the cut--off scale of our model, but it may be indicative of a
Nambu--Jona-Lasinio
result when the $Q\overline{Q}$ system is studied.
In fact, the fundamental issues raised by de Rafael and Taron
can in principle be explored in this scheme [18].  We will
defer this discussion to another place.

We believe there remains much to
do in  dynamical analyses of this kind
for heavy--light systems and their interactions.
 Our
model has inherent shortcomings. While the agreement of this crude model with
observation is
marginal at best, it suggests that improvements, such as a Pagels--Stokar
approximation, Holdom's  approach [22, 23], or
``Russian sum--rule'' methods [24], will lead to more
reliable estimates of crucial heavy meson observables.
The singular behavior
of the wave--function is not expected in a more realistic
scheme.   Replacing our
pure $s$--wave dynamics by QCD ladder approximation is clearly of some
interest.
For example, pinning down a prediction
of $g_A$ or the Isgur--Wise function from such
models would be quite interesting.  The full range of
phenomenological applications of generalized models would seem to be
an interesting direction for future research.  This toy
scheme is a first step in that direction and highlights the
challenges and advantages for more elaborate approaches.

\vskip .1in
\noindent
{\bf Acknowledgements}
\vskip .1in
\noindent
We thank E. Eichten, J. Rosner and C. Werner for useful discussions.

\newpage
\noindent
{\bf Appendix A:  Fermion Loop Approximation}
\vskip .1in
\noindent
{\em (i) Zero momentum pions}
\vskip .1in

Let us now integrate out the heavy and light fermion fields in
eq.(12) to produce an effective Lagrangian for $B$ and $B'$.
This can be viewed as a block--spin renormalization of the
theory of eq.(12) defined at the scale $\mu=\Lambda$, to a new
scale $\mu < \Lambda$, and is analogous to the treatment
of the light quark dynamics in Appendix B.
We begin in the approximation of treating the $\sigma$ and $\pi$
fields as zero--momentum (constant in spacetime) backgrounds
(small momentum $\pi$ amplitudes are considered subsequently).
We note that the fermion propagators
take the form:
\beq
S_{HQ}(k) = \frac{i}{v\cdot k}\left(\frac{1+\slash{v}}{2}\right);
\qquad
S_{LQ}(k) =i\left(\frac{\slash{k}  + m_q + g\Sigma^5 }{k^2 - \Omega^2}
\right)
\eeq
where:
\beq
\Omega^2 = (m_q + {g}\sigma/2)^2 +
{g}^2\pi^2/4; \qquad \pi^2=\pi^a\pi^a;
\qquad
\Sigma^5 = \half\sigma + i\half \gamma_5\pi\cdot \tau
\eeq
We obtain
from the diagram of Fig.(1)
(recall that the ${\cal{B}}$ contain $(1+\slash{v})/{2}$
projection factors):
\bea
i{\cal{S}}_{BB} & = &
-{g^2N} \int \frac{d^4 k}{(2\pi)^4}
\nonumber \\
& & \Tr \left[ ( -i\overline{\cal{B}'}\gamma^5
+ \overline{\cal{B}}  )
 \left(\frac{\slash{k} - \slash{p} + m_q + g\Sigma^5 }{(k - p)^2 - \Omega^2}
\right) ( -i\gamma^5{\cal{B}'}
+ {\cal{B}}  )
 \left(\frac{1+\slash{v}}{2}\right)\right] \frac{1}{v\cdot k}
\nonumber \\
& = &  -{g^2N} \Tr\left[( -i\overline{\cal{B}'}\gamma^5
+ \overline{\cal{B}}  )
I_1
( -i\gamma^5{\cal{B}'}
+ {\cal{B}}  )
\right]
\eea
and:
\beq
I_1 = \int \frac{d^4 k}{(2\pi)^4}
\left(\frac{(\slash{k} -\slash{p}) + m_q + g\Sigma^5 }{(k - p)^2 - \Omega^2
}\right)
 \frac{1}{v\cdot k}
\eeq

We carry out a ``block--spin'' integration over
heavy and light quark modes between the scales
$\mu$ and $\Lambda$ in Euclidean momentum space.
The integrals are evaluated with a Euclidean
4--momentum cut--off:
\bea
& & \int \frac{d^4 k}{(2\pi)^4}
\frac{1}{((k - p)^2 - \Omega^2) v\cdot k}
=  \frac{i}{16\pi }\left( \Lambda - \mu\right)
+\;  \frac{2i}{16\pi^2}v\cdot p \left[\ln(\Lambda^2 /\mu^2 )
\right]
\nonumber \\
& &\int \frac{d^4 k}{(2\pi)^4}
\frac{1}{((k - p)^2 - \Omega^2)}
=  -\frac{i}{16\pi^2 }\left[ \Lambda^2 - \mu^2 -
\Omega^2 \ln(\Lambda^2 /\mu^2)\right]
\nonumber \\
& & \int \frac{d^4 k}{(2\pi)^4} \frac{k_\mu}{(k^2 - \Omega^2) v\cdot k}
 =
-i\frac{v_\mu }{16\pi^2}\left[ \Lambda^2 - \mu^2 -
\Omega^2 \ln(\Lambda^2 /\mu^2)\right]
\nonumber \\
& & \int \frac{d^4 k}{(2\pi)^4} \frac{k_\mu}{(k^2 - \Omega^2)^2 v\cdot k}
 =
i\frac{v_\mu }{16\pi^2}\left[\ln(\Lambda^2/\mu^2 )\right]
\eea
Then $\Tr(I_1)$  can be written as
(note $\Tr(\overline{\cal{B}}\gamma^5{\cal{B}})
= \Tr(\overline{\cal{B}}'{\cal{B}}) = 0 $, etc.):
\bea
 & &i{g^2N}\Tr(\overline{\cal{B}}I_1{\cal{B}})
\nonumber \\
& &\qquad = -\half\Tr(\overline{\cal{B}}{\cal{B}})[( v\cdot p  +g\tilde\sigma
/2 )(Z_1 +4 Z_2 v\cdot p)+ Z_1(\Lambda+\mu)/\pi
- 2|\Omega|^2 Z_2]
\nonumber \\
& & i{g^2N}\Tr(\overline{\cal{B}}'(-i\gamma^5) I_1
(-i\gamma^5 ){\cal{B}}')
\nonumber \\
& & \qquad = -\half\Tr(\overline{\cal{B}}'{\cal{B}}')[( v\cdot p  -
g\tilde\sigma /2 )(Z_1 +4 Z_2 v\cdot p)+ Z_1(\Lambda+\mu)/\pi
- 2|\Omega|^2 Z_2 ]
\nonumber \\
& & i{g^2N}[\Tr(\overline{\cal{B}}'(-i\gamma^5) I_1
{\cal{B}}) + \Tr(\overline{\cal{B}} I_1
(-i\gamma^5 ){\cal{B}}')]
\nonumber \\
& & \qquad = -\half\Tr[\overline{\cal{B}}'
(g \pi\cdot \tau/2 ) {\cal{B}}
+
\overline{\cal{B}}
(g \pi\cdot \tau/2 ) {\cal{B}}']
(Z_1 +4 Z_2 v\cdot p)
\eea
where we let $g\widetilde{\sigma} = g\sigma + 2m_q$ and:
\bea
Z_1 & = &
\frac{g^2N}{8\pi }( \Lambda  - \mu ); \qquad
Z_2  = \frac{g^2N}{16\pi^2}\left[\ln(\Lambda^2/\mu^2)\right]
\eea
(note that the expression for $Z_1$  contains
a factor of $1/\pi$, not $1/\pi^2$).

\vskip 0.05in
\noindent
{\em (ii) The $g_A$ term}
\vskip 0.05in

Now consider small, but nonzero $(\sigma, \pi)$ momentum $q_\mu$,
with $q^2\approx 0$.
We compute the effective Lagrangian, where the $(\sigma, \pi)$ are
coupled through $\Sigma^5$. We then have
the amplitude of Fig.(2):
\bea
i{\cal{S}}_{BB\Sigma} & = &
{\half g^3 N} \int \frac{d^4 k}{(2\pi)^4}
\nonumber \\
&  &
\Tr \left[( -i\overline{\cal{B}'}\gamma^5
+ \overline{\cal{B}}  )
\frac{(\slash{k}+\slash{q}/2) (\sigma + i\pi\cdot\tau\gamma^5) (\slash{k}
-\slash{q}/2 )   }{((k +q/2)^2 - \Omega^2)((k- q/2)^2 - \Omega^2)v\cdot (k-p)}
( -i\gamma^5{\cal{B}'}
+ {\cal{B}}  )
\right]
\nonumber \\
& &
\eea
 We are interested
in the divergent terms of order $q$, since the $q=0$ term has previously
been computed:
\bea
& \approx & {-\frac{1}{4}g^3N} \int \frac{d^4 k}{(2\pi)^4}
\Tr \left[( -i\overline{\cal{B}'}\gamma^5
+ \overline{\cal{B}}  )
\frac{[\slash{k},\slash{q}] (\sigma - i\pi\cdot\tau\gamma^5) }{(k)^2
(k)^2 (v\cdot k)}
( -i\gamma^5{\cal{B}'}
+ {\cal{B}}  )
\right]
\nonumber \\
& = & {-\frac{i}{4}g^3N}\frac{1}{16\pi^2}
\Tr \left[( -i\overline{\cal{B}'}\gamma^5
+ \overline{\cal{B}}  )
{[\slash{v},\slash{q}] (\sigma - i\pi\cdot\tau\gamma^5) }
( -i\gamma^5{\cal{B}'}
+ {\cal{B}}  )
\right] \ln (\Lambda^2/\mu^2)
\nonumber \\
& = & {-\frac{i}{4}g Z_2}
\Tr \left[( -i\overline{\cal{B}'}\gamma^5
+ \overline{\cal{B}}  )
{[\slash{v},\slash{q}] (\sigma - i\pi\cdot\tau\gamma^5) }
( -i\gamma^5{\cal{B}'}
+ {\cal{B}}  )
\right]
\eea
If we now expand the result of eq.(65) we observe some
simplifications, e.g. $\overline{{\cal{B}}} [\slash{v},\slash{q}]
{\cal{B}} = 0$, and we obtain:
\bea
& = & {\frac{i}{2}g Z_2}
\Tr \left[
\overline{\cal{B}}
\slash{q}(  i\pi\cdot\tau\gamma^5 ) {\cal{B}}
- \overline{\cal{B}'}
\slash{q} ( i\pi\cdot\tau\gamma^5) {\cal{B}'}
 +i\overline{\cal{B}'}\gamma^5
\slash{q} \sigma
{\cal{B}}
+ \overline{\cal{B}}
i\gamma^5\slash{q} \sigma
{\cal{B}'}
\right]
\nonumber \\
\eea
This implies an operator in the effective Lagrangian of
the form:
\bea
& = & {\frac{1}{2}g Z_2}
\left(\Tr \left[
\overline{\cal{B}}\gamma^5
\gamma_\mu\tau^a {\cal{B}}
- \overline{\cal{B}'} \gamma^5
\gamma_\mu\tau^a  {\cal{B}'} \right]\partial^\mu\pi^a
- \Tr \left[ \overline{\cal{B}'}\gamma^5
\gamma_\mu  {\cal{B}}
+\overline{\cal{B}}\gamma^5
\gamma_\mu {\cal{B}'}
\right]\partial^\mu\sigma\right)
\nonumber \\
\eea

\newpage
\noindent
{\em (iii) Normalization Conventions}
\vskip .1in

Consider a complex scalar field $\Phi$ with the Lagrangian:
\beq
\partial_\mu\Phi^\dagger \partial^\mu\Phi - (M+\delta M)^2
\Phi^\dagger \Phi
\eeq
Define $\Phi' = \sqrt{2M}\exp(iMv\cdot x)\Phi$
and the Lagrangian becomes to order $1/M$:
\beq
 iv_\mu\Phi'{}^\dagger \partial^\mu\Phi' - \delta M
\Phi'{}^\dagger \Phi'
\eeq
 Now let $\widetilde\Phi =
\half(1+\slash{v})i\gamma^5\Phi'$ and write in terms of traces:
\beq
-i\half \Tr(\widetilde{\Phi}^\dagger v\cdot \partial \widetilde{\Phi})
+ \delta M\half \Tr(\widetilde{\Phi}^\dagger \widetilde{\Phi})
\eeq
Thus when the Lagrangian is
written in terms of ${\cal{B}}$ and ${\cal{B}}'$
the normal sign conventions are those of the vector mesons,
and opposite those of scalars, \ie,
the term in the Lagrangian $+\half \delta M \Tr({\cal\overline{B}B})$  an {\em
increase} in the
$B^5$ mass by an amount $ \delta M$. A properly normalized kinetic term is
$-i\half\Tr({\cal\overline{B}}v\cdot \partial{\cal{B}})$,
with the overall minus sign and $\half$.

\vskip 0.05in

\vskip 0.05in
\noindent
{\em (iv) Structure of the Effective Lagrangian}
\vskip 0.05in

The heavy meson effective Lagrangian therefore takes the form:
\bea
& & {\cal{L}}_{LH} = \nonumber \\
& & -i\half Z_1\Tr(\overline{{\cal{B}}}v\cdot\partial{\cal{B}})
-i\half Z_1\Tr(\overline{{\cal{B}}'}v \cdot \partial {\cal{B}}')
+ 2Z_2\Tr(\overline{{\cal{B}}}(v\cdot\partial)^2{\cal{B}})
+ 2Z_2\Tr(\overline{{\cal{B}}'}(v\cdot\partial)^2 {\cal{B}}')
\nonumber \\
& & -\frac{gZ_1}{4}
\left[\Tr(\overline{{\cal{B}}}\widetilde{\sigma} {\cal{B}})
 -
 \Tr(\overline{{\cal{B}}'}\widetilde{\sigma}{\cal{B}}')\right]
 -i{g Z_2}
\left[\Tr(\overline{{\cal{B}}}\widetilde{\sigma} v \cdot \partial{\cal{B}})
- \Tr(\overline{{\cal{B}}'}\widetilde{\sigma}v \cdot \partial{\cal{B}}')
\right]
\nonumber \\
& &- \frac{gZ_1 }{4} \left[
\Tr(\overline{{\cal{B}}'} \pi\cdot\tau  {\cal{B}}) + \Tr(\overline{{\cal{B}}}
\pi\cdot\tau {\cal{B}}')\right]
- i{gZ_2 }\left[
\Tr(\overline{{\cal{B}}'} \pi\cdot\tau v \cdot \partial {\cal{B}}) +
\Tr(\overline{{\cal{B}}}  \pi\cdot\tau v \cdot \partial{\cal{B}}')\right]
\nonumber \\
& &
+\frac{g^2Z_2}{4}
\left[\Tr(\overline{{\cal{B}}} (\widetilde{\sigma}^2 + \pi^2) {\cal{B}})
+ \Tr(\overline{{\cal{B}}'} (\widetilde{\sigma}^2  + \pi^2) {\cal{B}}')
\right]
\nonumber \\
& &
+ \;(\Lambda^2 - Z_1(\Lambda+\mu)/2\pi  )
\left[\Tr(\overline{{\cal{B}}}{\cal{B}}) +
\Tr({\overline{{\cal{B}}'}}{{\cal{B}}'})
\right]
\nonumber \\
&  &
+\frac{Z_2}{2} g
\Tr \left[
\overline{\cal{B}}\gamma^5
\slash{\partial}(\pi\cdot\tau ) {\cal{B}}
- \overline{\cal{B}'} \gamma^5
\slash{\partial} ( \pi\cdot\tau) {\cal{B}'}
- \overline{\cal{B}'}\gamma^5
\slash{\partial}(\sigma)   {\cal{B}}
- \overline{\cal{B}}
\gamma^5\slash{\partial}(\sigma) {\cal{B}'}
\right]
\eea
If we define:
\bea
{\cal{T}}& = &[1 + (4Z_2/Z_1)iv\cdot\partial]^{-1/2}
\eea
then eq.(71) becomes more compactly:
\bea
{\cal{L}}_{LH} & = & -i\half Z_1\Tr(\overline{{\cal{T}}{\cal{B}}}v \cdot
\partial
{\cal{T}} {\cal{B}})
  -i\half Z_1\Tr(\overline{{\cal{T}}{\cal{B}}'}v \cdot \partial
{\cal{T}} {\cal{B}}')
\nonumber \\
& & -\frac{g Z_1}{4}\left[
\Tr(\overline{{\cal{T}}{\cal{B}}}\widetilde{\sigma}{\cal{T}} {\cal{B}})
-\Tr(\overline{{\cal{T}}{\cal{B}}'}\widetilde{\sigma}{\cal{T}}{\cal{B}}')\right]
\nonumber \\
& &- \frac{gZ_1 }{4} \left[
\Tr(\overline{{\cal{T}}{\cal{B}}'} \pi\cdot\tau {\cal{T}} {\cal{B}}) +
\Tr(\overline{{\cal{T}}{\cal{B}}}  \pi\cdot\tau {\cal{T}}{\cal{B}}')\right]
\nonumber \\
& &
+\frac{g^2Z_2}{4}
\left[ \Tr(\overline{{\cal{B}}} (\widetilde{\sigma}^2 + \pi^2) {\cal{B}})
+ \Tr(\overline{{\cal{B}}'} (\widetilde{\sigma}^2  + \pi^2) {\cal{B}}')\right]
\nonumber \\
& &
+\;(\Lambda^2 - Z_1(\Lambda+\mu)/2\pi  )
\left[\Tr(\overline{{\cal{B}}}{\cal{B}}) +
\Tr({\overline{{\cal{B}}'}}{{\cal{B}}'}]
\right)
\nonumber \\
&  &
+\frac{Z_2}{2} g
\Tr \left[
\overline{\cal{B}}\gamma^5
\slash{\partial}(\pi\cdot\tau ) {\cal{B}}
- \overline{\cal{B}'} \gamma^5
\slash{\partial} ( \pi\cdot\tau) {\cal{B}'}
- \overline{\cal{B}'}\gamma^5
\slash{\partial}(\sigma)   {\cal{B}}
- \overline{\cal{B}}
\gamma^5\slash{\partial}(\sigma) {\cal{B}'}
\right]
\nonumber \\
\eea
To simplify the subsequent analysis we will assume that the
subleading terms of order $Z_2 v\cdot p /Z_1 $ are negligible, and take
${\cal{T}}=1$.  Since these terms arise upon expanding
the loop integrals in powers of $1/\Lambda$, we cannot self--consistently
use the effective Lagrangian in this form
unless this condition is
at least approximately valid. We see that other terms,
such as the last one in eq.(73) which leads to $g_A$, are leading in this order
and describe  various physical processes. Thus, we expect
the amplitudes these terms describe to be small.  If $4Z_2 v\cdot p/Z_1$
is large, then we must retain full analytic expressions for the loop
integrals to fit the theory.

We see that there is thus an induced kinetic term for the ${\cal{B}}$
and ${\cal{B}}'$ fields with a common wave--function normalization.
We absorb the factor $Z_1$ into the
fields as ${\cal{B}}\rightarrow \sqrt{Z_1 }^{-1} {\cal{B}}$.
Thus, with the field  redefinition
we then have the full effective Lagrangian:
\bea
{\cal{L}}_{LH} & = & -i\half \Tr(\overline{{\cal{B}}}v \cdot \partial
{\cal{B}})
  -i\half \Tr(\overline{{\cal{B}}'}v \cdot \partial
{\cal{B}}')
\nonumber \\
& & -\frac{g }{4}\left[  \Tr(\overline{{\cal{B}}}\widetilde{\sigma} {\cal{B}})
 -\Tr(\overline{{\cal{B}}'}\widetilde{\sigma}{\cal{B}}')\right]
- \frac{g }{4} \left[
\Tr(\overline{{\cal{B}}'} \pi\cdot\tau  {\cal{B}}) + \Tr(\overline{{\cal{B}}}
\pi\cdot\tau {\cal{B}}')\right]
\nonumber \\
& &
+\frac{g^2Z_2}{4Z_1}
\left[ \Tr(\overline{{\cal{B}}} (\widetilde{\sigma}^2 + \pi^2) {\cal{B}})
+ \Tr(\overline{{\cal{B}}'} (\widetilde{\sigma}^2  + \pi^2) {\cal{B}}')\right]
\nonumber \\
& &
+\;\Delta \left[\Tr(\overline{{\cal{B}}}
{\cal{B}}) + \Tr({\overline{{\cal{B}}'}}
{{\cal{B}}'})
\right]
\nonumber \\
&  &
+\frac{Z_2}{2Z_1} g
\Tr \left[
\overline{\cal{B}}\gamma^5
\slash{\partial}(\pi\cdot\tau ) {\cal{B}}
- \overline{\cal{B}'} \gamma^5
\slash{\partial} ( \pi\cdot\tau) {\cal{B}'}
- \overline{\cal{B}'}\gamma^5
\slash{\partial}(\sigma)   {\cal{B}}
- \overline{\cal{B}}
\gamma^5\slash{\partial}(\sigma) {\cal{B}'}
\right]
\nonumber \\
\eea
where:
\beq
\Delta = \frac{1}{Z_1}(\Lambda^2 - Z_1(\Lambda+\mu)/2\pi  )
\eeq
The equation of motion in momentum space is $v\cdot p = 2\Delta
+ ...$ and $2\Delta $  is the mass difference between the
heavy meson and the heavy quark in the chiral symmetric phase:
\beq
M_B = 2\Delta + M_Q
\eeq
Note that $\Delta > 0$ ($\Delta < 0$) for $g^2N/16\pi^2 < 1$
 ($g^2N/16\pi^2 > 1$).

\vskip .05in
\noindent
{\bf Appendix B:  Light Quark Dynamics }
\vskip .05in
\noindent
{\em (i) Deriving the Constituent Quark Model }
\vskip .1in

The effective Lagrangian in the light quark sector is:
\beq
{\cal{L}} = \overline{\psi}(i\slash{\partial} - m_q )\psi
-\frac{g^2}{\Lambda^2}\overline{\psi}
\gamma_\mu\frac{\lambda^A}{2}\psi\overline{\psi}
\gamma^\mu\frac{\lambda^A}{2}\psi
\eeq
For concreteness we will take $\psi = (u, d)$, and in
the limit that the quark mass matrix $m_q\rightarrow 0$, we
have an exact chiral $SU(2)\times SU(2)$ invariant
Lagrangian. This can be viewed as a single  gluon exchange potential,
where we assume a
``gluon mass'' $\Lambda/\sqrt{2}$, and we have written the form
of the effective Lagrangian at  $q^2 \sim \Lambda^2$, integrating out
the massive gluon, and truncating on $dim=6$
operators.

Upon Fierz--rearrangment of the interaction Lagrangian,
keeping only leading terms in
$1/N_{C}$, eq.(77) takes the form:
\bea
 {\cal{L}}_L  & = & \overline{\psi}(i\slash{\partial} - m_q )\psi
+
\frac{g^2}{\Lambda^2}\left(
\overline{\psi}_L\psi_{R}\overline{\psi}_{R}\psi_{L} +
\overline{\psi}_L\tau^A\psi_{R}\overline{\psi}_{R}\tau^A\psi_{L}\right.
\nonumber \\
& &
-\frac{1}{8}\overline{\psi}\gamma_\mu\tau^A\psi
\overline{\psi}\gamma^\mu\tau^A\psi
-\frac{1}{8}\overline{\psi}\gamma_\mu\gamma_5\tau^A\psi
\overline{\psi}\gamma^\mu\gamma_5\tau^A\psi
\nonumber \\
& & \left.
-\frac{1}{8}\overline{\psi}\gamma_\mu\psi \overline{\psi}\gamma^\mu\psi
-\frac{1}{8}\overline{\psi}\gamma_\mu\gamma_5\psi
\overline{\psi}\gamma^\mu\gamma_5\psi \right)
\eea
where $\psi_L = (1-\gamma_5)\psi/2$,
$\psi_R = (1+\gamma_5)\psi/2$.
Here $\tau^A$ are Pauli matrices acting upon the isospin indices.

For the present analysis we will truncate eq.(78) on the pure
Nambu--Jona--Lasinio terms, since the (vector)$^2$ and
(axial-vector)$^2$ terms play no significant role in the chiral dynamics
(they are associated with the formation of virtual $\rho$ and
$A_1$ vector mesons in the model). Hence we take:
\bea
 {\cal{L}}_L  & = & \overline{\psi}(i\slash{\partial} - m_q )\psi
+
\frac{g^2}{\Lambda^2}\left(
\overline{\psi}_L\psi_{R}\overline{\psi}_{R}\psi_{L} +
\overline{\psi}_L\tau^A\psi_{R}\overline{\psi}_{R}\tau^A\psi_{L}\right)
\eea
We can solve the light--quark dynamics in large--$N$
in the usual way  by writing an
equivalent effective Lagrangian of the form:
\bea
{\cal{L}} & = & \overline{\psi}(i\slash{\partial} - m_q )\psi
- g\overline{\psi}_L\Sigma\psi_R - g\overline{\psi}_R\Sigma^\dagger\psi_L
- \half\Lambda^2\Tr(\Sigma^\dagger\Sigma)
\eea
where:
\beq
 \Sigma = \half\sigma I_2 + i\pi^a\frac{\tau^a}{2}
\eeq
is an auxiliary  field.  We emphasize that at this stage $\Sigma$
is a $2\times 2$
complex field, so both
$\sigma$ and $\pi^a$ are complex,  (otherwise, with $\sigma$ and $\pi$ real
there would be
unwanted contributions from $\VEV{T \Sigma\;\Sigma}
= \VEV{T \Sigma^\dagger \;\Sigma^\dagger } \neq 0$ in integrating
out $\Sigma$).

 Thus there is parity doubling at this stage,
$Im(\sigma)$ is the
fourth Goldstone boson associated with the $U(1)$ problem,
and $Im(\pi^a)$ is the $0^+$ isotriplet.
 The restriction to real $\pi^a$ will emerge
dynamically at very low energies,  since the induced
$\Tr(\Sigma^\dagger \Sigma \Sigma^\dagger \Sigma)$ term
will lift the degeneracy of the $Re(\pi)$ and $Im(\pi)$.
We ultimately must add a $\det(\Sigma) + h.c.$ term to get
rid of the  $Im(\sigma)$ mode.

We now integrate out the fermion fields
on scales $\Lambda^2 > q^2
> \mu^2$, keeping only the leading large--$N_C$ fermion
loop contributions.  We use the
massless fermion propagator, treating
$\Sigma$ as a classical background field.
Thus we arrive at an effective field theory at the
scale $\mu$:
\bea
{\cal{L}} & = & \overline{\psi}(i\slash{\partial} - m_q )\psi
- g\overline{\psi}_L\Sigma\psi_R - g\overline{\psi}_R\Sigma^\dagger\psi_L
\nonumber \\
& & + Z_2 \Tr(\partial_\mu\Sigma^\dagger\partial^\mu\Sigma) - V(\Sigma)
\eea
where:
\bea
Z_2 &=& \frac{g^2N}{16\pi^2}\ln(\Lambda^2/\mu^2)
\nonumber \\
V(\Sigma) &= & \left[\half\Lambda^2 - \frac{g^2N}{8\pi^2}(\Lambda^2 - \mu^2)
\right]\Tr(\Sigma^\dagger\Sigma) -
 \frac{gN}{8\pi^2}(\Lambda^2 - \mu^2)\Tr (m_q\Sigma + h.c.)
\nonumber \\
& & +  \frac{g^4N}{16\pi^2}\ln(\Lambda^2/\mu^2)\Tr(\Sigma^\dagger\Sigma
\Sigma^\dagger\Sigma)
\eea
We see that $Z_2\rightarrow 0$ as $\mu\rightarrow \Lambda$,
reflecting the compositeness of the $\Sigma$ field.
Let us now renormalize the $\Sigma$ field:
\beq
\Sigma\rightarrow \sqrt{Z_2}\Sigma
\eeq
and we have the properly normalized effective Lagrangian at
the scale $\mu$ (this is proper normalization for
real $\sigma$ and $\pi$):
\bea
{\cal{L}} & = & \overline{\psi}(i\slash{\partial} - m_q )\psi
- \tilde{g}\overline{\psi}_L\Sigma\psi_R -
\tilde{g}\overline{\psi}_R\Sigma^\dagger\psi_L
\nonumber \\
& & + \Tr(\partial_\mu\Sigma^\dagger\partial^\mu\Sigma) - \tilde{V}(\Sigma)
\eea
where:
\bea
\tilde{g} &\equiv & 1/\sqrt{Z_2}
\nonumber \\
\tilde{V}(\Sigma) &= & m_\sigma^2\Tr(\Sigma^\dagger\Sigma) -
\omega\Tr (m_q\Sigma + h.c.)
\nonumber \\
& & +  \lambda \Tr(\Sigma^\dagger\Sigma
\Sigma^\dagger\Sigma)
\nonumber \\
m^2_\sigma & = &
\left(\frac{1}{Z_2}\right)
\left[\half \Lambda^2 - \frac{g^2N}{8\pi^2}(\Lambda^2 - \mu^2)
\right]
\nonumber \\
\lambda & = & \frac{16\pi^2}{N\ln(\Lambda^2/\mu^2} = \tilde{g}^2
\nonumber \\
\omega & = &  \frac{\tilde{g}gN}{8\pi^2  }(\Lambda^2 - \mu^2)
\eea
The effective Lagrangian is seen to be a linear $\sigma$--model
at scales $\mu < \Lambda$.
As the scale $\mu\rightarrow 0$
we see that the theory is trivial, since $\tilde{g}\rightarrow 0$.
However, these evolution results apply only
to a scale $\mu_0$ corresponding to a mass scale
for the fermion. Nonzero $m_q$ will block the evolution
into the far infrared, but we will neglect this presently.
The theory will develop a chiral instability
(a constituent quark mass) provided that $m_\sigma^2$ becomes
tachyonic (negative) at some scale $\mu_0$.
By tuning the bare coupling constant $g^2$ we can put the
model in a symmetric phase, $m^2>0$ $\rightarrow$ ${g^2N}/{4\pi^2}<1$,
or
in a chiral symmetry breaking phase: $m^2<0$ $\rightarrow$ ${g^2N}/{4\pi^2}>1$,
where the critical bare coupling corresponds
to $m_\sigma^2=0 $ as $\mu_0\rightarrow 0$.

In the broken phase (ignoring $m_q$)
the $\sigma$ field develops a vacuum
expectation value $\VEV{\sigma} = f_\pi =
\sqrt{2}|m_\sigma |/\sqrt{\lambda}$. We see that
the renormalized $\sigma$ field develops a vacuum
expectation value given by:
\beq
\VEV{\sigma}_r^2  = Z_2 \left(\frac{16\pi^2\Lambda^2}{g^4N
\ln(\Lambda^2/\mu^2)}
\right)
\left( \frac{g^2N}{4\pi^2} - 1\right)
= \left(\frac{\Lambda^2}{g^2 }
\right)
\left( \frac{g^2N}{4\pi^2} - 1\right)
\eeq
 In the broken phase we can then write
$\sigma= f_\pi + \hat{\sigma}$, and
the physical mass$^2$ of the $\hat{\sigma}$ is readily
seen to be $m_{\hat{\sigma}}^2 =2 |m^2_\sigma|$, while the fermion mass becomes
$m_0=\half f_\pi \tilde{g}$.  Thus, using eqs.(86)
to relate $\tilde{g}^2=\lambda$,
we obtain  the usual
Nambu--Jona--Lasinio result: $m_{\tilde{\sigma}} = 2m_0$.

The solution to the theory can thus be written as a chiral quark model
in which we have both constituent quarks described by $\psi$ and
the mesons decribed by $\Sigma$.  In the broken phase
it is useful to pass to a nonlinear $\sigma$--model and write:
\beq
\Sigma \rightarrow \half f_\pi \exp(i\pi^a\tau^a/f_\pi)
\eeq
and:
\bea
{\cal{L}} & = & \overline{\psi}(i\slash{\partial} - m_q )\psi
- m_0\overline{\psi}_L\exp(i\pi^a\tau^a/f_\pi)\psi_R - m_0
\overline{\psi}_R \exp(-i\pi^a\tau^a/f_\pi) \psi_L
\nonumber \\
& & + \Tr(\partial_\mu\Sigma^\dagger\partial^\mu\Sigma)
+ \omega\Tr (m_q\Sigma + h.c.)
\eea
where $m_0=\half \tilde{g}f_\pi$ is the constituent quark mass.
Note, in our present normalization conventions that $f_\pi =
93$ MeV.  By a chiral redefinition of the fields,
$\psi_R\rightarrow \xi\psi_R$ and $\psi_L\rightarrow \xi^\dagger\psi_L$
we arrive at the Georgi--Manohar Lagrangian (their eq.(2.9)) with
$g_A = 1.0$ (note that they fit $G_A/G_V = (5/3)g_A$ and obtain
$g_A= 0.75$, consistent with our large--N approximation).

When the $\sigma$ and $\pi$ fields
are slowly varying in space, the light quark propagator
of the chiral quark model  is given by
(in terms of the unrenormalized fields):
\bea
S_F & = & i\left( \slash{p} - m_q - {g}\half\sigma - i{g}\half \gamma_5\pi\cdot
\tau
  \right)^{-1}
\nonumber \\
& = & i\left(
\frac{\slash{p} + m_q + {g}\Sigma^5}{
p^2 - \Omega^2}
  \right)
\eea
where we  define:
\beq
\Omega^2 = (m_q + {g}\sigma/2)^2 +
{g}^2\pi^2/4; \qquad \pi^2=\pi^a\pi^a
\eeq
\beq
\Sigma^5 = \half\sigma + i\half \gamma_5\pi\cdot \tau
\eeq
In the broken phase we replace $\sigma = f_\pi$ and $\Sigma
\rightarrow \half f_\pi \exp(i\pi^a\tau^a/f_\pi)$.
For future ease of writing we can often replace
${g}\widetilde{\sigma}/2 = {g}\sigma/2 + m_q\sqrt{Z_2}$
since it easy to restore the explicit chiral symmetry breaking quark mass
terms.

\vskip .05in
\noindent
{\em (ii) Schematic Discussion of a Parity Doubled Nucleon }
\vskip .1in

Consider a ``nucleon'' doublet $N$ with the $SU(2)_L\times SU(2)_R$
assignments $N_L\sim (\half,0)$, $N_R\sim (0,\half)$.  Also,
we introduce a  partner, $K$, of opposite parity with assignments
$K_L\sim (0, \half)$, $K_R\sim (\half,0)$.
A typical renormalizeable linear $\sigma$--model
effective matter Lagrangian
(not including the $\Sigma$ kinetic and potential terms) is then:
\bea
{\cal{L}} & =  & \overline{N}i\slash{\partial}N +
\overline{K}i\slash{\partial}K
\nonumber \\
& & - M_1 \overline{N}_L \Sigma N_R - M_2  \overline{K}_L \Sigma^\dagger K_R
- M_0  \overline{N}_L K_R - M'_0  \overline{N}_R K_L + h.c.
\eea
Parity symmmetry requires  $M_0=M_0'$.  We
consider the special case $M_1=M_2=M$, which is the analogue of our
model, but this is not generally required by symmetries.  Now perform the
redefinitions, $N_L\rightarrow \xi N_L$, $K_L\rightarrow \xi^\dagger K_L$,
$N_R\rightarrow \xi^\dagger N_R$, $K_R\rightarrow \xi K_R$.
Thus, the Lagrangian becomes:
\bea
{\cal{L}} & =  & \overline{N}(i\slash{\partial} + {\cal\slash{V}}
+ \gamma^5{\cal\slash{A}})N + \overline{K}(i\slash{\partial} + {\cal\slash{V}}
- \gamma^5{\cal\slash{A}})K
\nonumber \\
& & - M \overline{N}N - M  \overline{K} K
- M_0  \overline{N} K - M_0  \overline{K} N + h.c.
\eea
Upon diagonalizing, the
mass eigenfields are just $(N \pm K)/\sqrt{2}$, with mass eigenvalues
$M\pm M_0$.  We can decouple the heavier state by setting
$(N + K)/\sqrt{2} = 0$, whence the light effective Lagrangian
for $Q=(N - K)/\sqrt{2}$ is:
\bea
{\cal{L}} & =  & \overline{Q}(i\slash{\partial} + {\cal\slash{V}}
)Q - (M-M_0) \overline{Q}Q
\eea
We see that $g_A=0$.  Hence,
$g_A$ is not generally of order
unity as is the case of a non--parity doubled nucleon.
(this is also a consequence of the special
case  $M_1 = M_2$; more generally $g_A=\sin(2\theta)$
where $\theta$ is the mass mixing angle). With $g=0$ the only nontrivial
Goldberger--Treiman relation refers to
the pionic transition amplitude between the ground state, $Q$,
and the parity partner.

\vskip .1in
\noindent
{\bf References}
\vskip .1in
\noindent
\begin{enumerate}

\item N. Isgur and M.B. Wise, Phys.\ Lett.\ B {\bf 232}, 113 (1989);
{\bf 237}, 527 (1990), Nucl.\ Phys.\ {\bf B348}, 276 (1991).

\item
E. Eichten and F. Feinberg, Phys.\ Rev.\ D {\bf 23}, 2724 (1981);
E. Eichten and B. Hill, Phys.\ Lett.\ B {\bf 234}, 511 (1990);
{\bf 243}, 427 (1990).

\item M. Suzuki, {\em Nucl. Phys.} {\bf B258} 553 (1985).

\item M.B. Voloshin and M.A. Shifman, Yad.\ Fiz.\ {\bf 45}, 463 (1987)
[Sov.\ J.\ Nucl.\ Phys.\ {\bf 45}, 292 (1987)]; {\bf 47}, 801 (1988)
[{\bf 47}, 511 (1988)].

\item
H. Georgi, {\em Nucl. Phys.} {\bf B348}, 293 (1991);
 {\em Phys. Lett.} {\bf B240}, 447 (1990).

\item M. B. Wise,  {\em Phys. Rev.} {\bf D45} 2188 (1992)

\item G. Burdman and J. Donoghue, {\em Phys. Lett.} {\bf B280}, 287 (1992)

\item T. M. Yan \etal,  {\em Phys. Rev.} {\bf D46} 1148 (1992)

\item  E. Jenkins, M. Savage, {\em  Phys. Lett.} B {\bf 281}, 331 (1992);
B. Grinstein,  E. Jenkins,  A. Manohar,
M. Savage, M. Wise, {\em Nucl. Phys.} {\bf B380} 369 (1992).

\item P. Cho, ``Heavy Hadron Chiral Perturbation Theory,''
Harvard Univ. Preprint, HUTP--92--A039, August (1992);
 Hai--Yang Cheng, \etal, ``Chiral Lagrangians for
Radiative Decays of Heavy Hadrons,'' Cornell Preprint,
CLNS--92--1158, Sept. (1992).

\item J. F. Amundson, \etal, {\em Phys. Lett.} {\bf B296}, 415 (1992).

\item A. Falk, ``Excited Heavy Mesons and Kaon Loops in Chiral
Perturbation Theory,'' SLAC--Pub--6055 (1993); A. Falk, M. Luke,
{\em Phys. Lett.} {\bf B292}, 119 (1992).

\item  A. Manohar, H. Georgi, {\em Nucl. Phys.} {\bf B234}, 189 (1984);
B. Holdom, J. Terning {\em Phys. Lett.} {\bf B245}, 612 (1990).

\item Ming--Lu, M. Wise, N. Isgur, {\em Phys. Rev. } {\bf D45}, 1553 (1992).

\item M. Luke and A. Manohar,  {\em Phys. Lett.} {\bf B286}, 348 (1992).

\item A. Falk, M. Neubert, and M. Luke,{\em  Nucl. Phys.} {\bf B388}, 363
(1992).

\item C. T. Hill and E. J. Eichten, ``Reparametrization Invariance and
Heavy Quarks'' Fermi--Pub--92/175-T (in preparation).

\item E. de Rafael, J. Taron  {\em Phys. Lett.} {\bf B282}, 215 (1992);
see however, B. Grinstein and P. Mende, ``On Constraints of Heavy--Meson
Form Factors,'' SSC Lab. preprint SSCL--PP--167, Nov. (1992).

\item
E. Eichten, K. Gottfried, T. Kinoshita, K. Lane, T. Yan,
{\em Phys. Rev.} {\bf D21},  203 (1980).

\item see, e.g., C. Bernard, J. Labrenz, A. Soni, {\em Nucl. Phys. B
(Proc. Suppl.)} {\bf 26}, 384 (1992).

\item J. Rosner, private communication.

\item H. Pagels, S. Stokar, {\em Phys. Rev.} {\bf D20},  2947 (1980);
B. Holdom, {\em Phys. Rev.} {\bf D45},  2534 (1992).

\item B. Holdom, M. Sutherland, ``Simply Modeling Meson HQET,'' UTPT--92--16
(Nov., 1992).

\item M. Shifman, private communication.

\end{enumerate}
\end{document}